\documentclass[a4paper,fleqn]{cas-dc}

\usepackage[numbers]{natbib}
\usepackage{float}
\usepackage{siunitx}
\DeclareSIUnit\angstrom{\text{Å}}
\usepackage{graphicx}
\usepackage{amssymb}
\usepackage{amsmath,amsthm,nccmath}
\usepackage{xcolor}
\usepackage{textcomp}

\usepackage{graphicx}

\usepackage{caption}
\usepackage{subcaption}
\usepackage{booktabs}
\usepackage{enumitem}
\usepackage{gensymb}

\usepackage{array}

\def\tsc#1{\csdef{#1}{\textsc{\lowercase{#1}}\xspace}}
\tsc{WGM}
\tsc{QE}

\begin{document}
\let\WriteBookmarks\relax
\def\floatpagepagefraction{1}
\def\textpagefraction{.001}

\shorttitle{}    

\shortauthors{}  

\title [mode = title]{Defect-controlled electrical and optical properties of CrN thin films: experiment and first-principles study}  



%

\author[1]{J. Bul\'i\v{r}}[orcid=0000-0002-1302-4365]
\cormark[1]


\ead{bulir@fzu.cz}

\ead[url]{https://www.fzu.cz/lide/dr-ing-jiri-bulir}


\cortext[1]{Corresponding author}

\fntext[1]{}

\author[2]{U. D. Wdowik}
\author[1]{J. More-Chevalier}
\author[1]{P. Hub\'ik}
\author[1]{E. de Prado}
\author[1]{M. Vondr\'a\v{c}ek}
\author[1]{L. Fekete}
\author[1]{M. Novotn\'{y}}
\author[1]{J. Lan\v{c}ok}
\author[2,3]{D. Legut}

\affiliation[1]{organization={Institute of Physics of the Czech Academy of Sciences},
            addressline={ Na Slovance 2}, 
            city={Prague 8},
            postcode={18200}, 
            state={},
            country={Czech Republic}}





\affiliation[2]{organization={IT4Innovations, VSB - Technical University of Ostrava},
            addressline={17. listopadu 2172/15}, 
            city={Ostrava-Poruba},
            postcode={70800}, 
            state={},
            country={Czech Republic}}

\affiliation[3]{organization={Department of Condensed Matter Physics, Faculty of Mathematics and Physics, Charles University},
            addressline={Ke Karlovu 3}, 
            city={Praha 2},
            postcode={12116}, 
            state={},
            country={Czech Republic}}


\begin{abstract}
Thin chromium nitride (CrN) films were deposited by RF magnetron sputtering in a reactive atmosphere of Ar-N $_2$ on fused silica and MgO (001) substrates at temperatures between 400 $\degree$C and 800 $\degree$C. The influence of nitrogen content and substrate temperature on structural, electrical, and optical properties was systematically investigated. The overstoichiometric CrN films exhibited a reduced resistivity below 6 m$\Omega \cdot$cm with stable p-type behavior, while increasing the substrate temperature above 600 $\degree$C induced a transition to n-type conductivity. Spectroscopic ellipsometry in the UV–Vis–NIR and infrared spectral ranges revealed a strong dependence of the dielectric function and the absorption edge on film stoichiometry and deposition conditions. Structural characterization using X-ray diffraction and atomic force microscopy confirmed substrate-dependent crystallinity and grain growth. Density functional theory calculations showed that cation and anion vacancies strongly modify the electronic structure and optical response, explaining the experimentally observed conductivity transitions. The results demonstrate that the deposition parameters provide an effective means to tailor CrN thin films via defect engineering.

\end{abstract}


\begin{highlights}
\item  CrN thin films deposited by reactive RF magnetron sputtering.
\item Nitrogen content affects electrical resistivity of CrN films.
\item Growth temperature modifies carrier type and electrical resistivity.
\item Optical constants determined by UV–Vis–IR spectroscopic ellipsometry.
\item Defect states explain conductivity changes from DFT calculations.
 
\end{highlights}


\maketitle


\section{Introduction}
Chromium nitride (CrN) is a transition-metal nitride that has attracted significant attention for its unique combination of mechanical, electrical, and optical properties, making it an ideal material for a wide range of applications, from protective coatings to semiconductor devices. CrN is a very hard material with excellent resistance to high-temperature oxidation, wear, and corrosion. Some transition-metal nitrides, based primarily on scandium nitride and chromium nitride, have seen increased interest for energy harvesting via thermoelectricity and piezoelectricity \cite{Eklund2016, Sabeer2021, Gharavi2021, More-Chevalier2025, More-Chevalier2026, quintela_epitaxial_2015}. Of these materials, CrN exhibits excellent thermoelectric properties with a large Seebeck coefficient and a high thermoelectric power factor \cite{Sabeer2021}. The versatility of CrN is underpinned by its diverse material characteristics, which are influenced by its stoichiometry, crystallinity, and microstructure. Understanding the optical and electrical properties of CrN is crucial for optimizing its performance in various applications, particularly in energy harvesting via the thermoelectric effect.

Its electrical conductivity is strongly influenced by factors such as nitrogen content, crystal structure, and temperature. Substoichiometric CrN exhibits electrical resistivity lower than that of pure chromium, a behavior closely linked to the magnetic properties of Cr. Specifically, the incorporation of a small amount of nitrogen suppresses antiferromagnetic coupling, thereby enhancing the charge-carrier mobility \cite{garzon-fontecha_role_2018}. As a result, substoichiometric CrN displays metallic-like behavior, with electrical conductivity dominated by free electrons. In contrast, in overstoichiometric CrN, the concentration of free electrons decreases, and electrical transport is increasingly governed by hole conduction, leading to a positive Hall coefficient and increased electrical resistivity. However, at higher nitrogen concentrations, a subsequent decrease in resistivity has been observed \cite{le_febvrier_p_2022}, indicating a complex dependence of charge transport on stoichiometry. CrN exhibits a narrow band gap whose reported value strongly depends on its stoichiometry and defect concentration. Values ranging from approximately 0.02 to 0.8 eV have been reported based on electrical-transport \cite{zhang_variable-range_2011, jin_strain-mediated_2021, dinh_carrier_2026} and optical-spectroscopy measurements \cite{alam_electronic_2022,zhang_crn_2010, gall_growth_2002}. These results highlight the strong sensitivity of the electronic structure and charge-transport behavior of CrN to its stoichiometry and defect structure.

The optical properties are closely related to their electrical behavior, as charge-carrier dynamics and photon absorption are interdependent. Optical characteristics such as absorption, reflectivity, and transmission are governed by the electronic structure and its interaction with electromagnetic radiation. Optical analysis of CrN coatings is studied using transmittance and reflectance measurements in a wide spectral range from 250 nm to 30 µm in the publication \cite{zhang_crn_2010}. The formation of CrN and Cr$_2$N phases was studied using spectral ellipsometry in the publication \cite{aouadi_spectroscopic_2001}. In the aforementioned publication, the authors consider the optical model as a mixture of two phases (Cr/Cr$_2$N or Cr$_2$N/CrN) and approximate it using the effective Bruggeman medium. Ellipsometry was also used to monitor the growth and formation of the aforementioned phases \cite{aouadi_control_2002}. CrN is a compound in which Cr/N stoichiometry is not strictly constrained. Consequently, its optical properties are highly sensitive to compositional variations, with changes in the N/Cr ratio significantly affecting the absorption edge and the overall spectral response.

The most widely used method for preparing CrN coatings is magnetron sputtering \cite{gall_growth_2002, garzon-fontecha_role_2018, barata_characterisation_2001, gharavi_synthesis_2019, gharavi_microstructure_2018}. The reported physical properties of CrN coatings vary significantly with deposition parameters. These include the substrate temperature and the $N_2$-Ar ratio in the working gas. The choice of substrate material influences the growth mode. The MgO substrate is often used for the epitaxial growth of CrN \cite{gall_growth_2002, aouadi_spectroscopic_2001}.

This paper aims to analyze the optical and electrical properties of CrN, focusing on its absorption spectrum (as represented by the dielectric functions) and its electrical conductivity. In particular, we focused on depositing CrN films with p-type conductivity, achieved by tuning the deposition conditions, primarily by increasing the N$_2$/Ar flow rate during deposition. By examining both experimental results and theoretical models, we seek to elucidate the mechanisms governing the interplay between the optical and electrical behavior of CrN. This dual approach will help uncover insights into the potential of CrN-based materials for advanced applications in optoelectronics, energy harvesting, and protective coatings.

\section{Methodology}
\subsection{Experimental}
\subsubsection{Sample preparation}

The CrN films were deposited in a vacuum chamber by reactive magnetron sputtering of a chromium target with a diameter of 100 mm. The vacuum chamber was evacuated to an ultimate pressure of $2\times10^{-5}$ Pa before the deposition process. An argon-nitrogen reactive atmosphere was prepared by mixing Ar and $N_2$ at a flow rate of 2 to 5 sccm and 8 to 18 sccm, respectively. The pressure was adjusted to 1 Pa by the throttle valve between the deposition chamber and the vacuum turbomolecular pump. A 200 W radio-frequency power (13.56 MHz) was applied to generate a plasma discharge. The Cr target was pre-sputtered for 120 seconds before the deposition to clean the target surface of the native oxide. We used a shutter in front of the sputtered target to prevent deposition during cleaning. For the substrate, we chose two materials: fused silica and MgO (001). The former is a common material suitable for both electrical and optical analysis. We used it to study CrN film grown in a polycrystalline structure. The latter material was used to study CrN with an oriented crystalline structure. The substrates were heated to temperatures between $400\degree$C and $800\degree$C during deposition. The MgO substrates were annealed at $1000\degree$C for 1 hour for crystalline surface reconstruction before deposition. Two series of experiments were performed: (i) a series of N$_2$ / Ar gas compositions and (ii) a series of substrate temperatures. The deposition times were adjusted to approximately 360 s for the first and 2000 s for the second, resulting in different CrN layer thicknesses. The deposition parameters are shown in Table \ref{tab:Fit_N2Ar}.

\begin{table*}
\small
\centering
\caption{Deposition conditions of CrN films including N$_2$/Ar gas flow ratios and substrate temperature. The table includes the CrN composition measured by XPS on the untreated sample and after Ar$^+$ sputtering, film thickness (Ellipsometry), RMS roughness (AFM), band-gap energy (Transmittance), and out-of-plane c lattice parameter (XRD).}
\begin{tabular}{cccccccc}
    \toprule
    T  & Gas & \multicolumn{2}{c}{N/Cr}  & Thickness & Roughness & Eg & c  $\pm 0.005$ \\
    ($\degree$C) & N$_2$/Ar & Untreated & Sputt. & (nm) & (nm) & (eV) & (\AA) \\
    \midrule
    450 & 1.6 & 1.0 & 0.77 & 37 & 0.8 & 0.74 & \\
    450 & 2.4 & 1.14 & 0.89 & 33 & 0.7 & 0.45 & \\
    450 & 3 & 1.03 & 0.96 & 37 & 0.6 & 0.34 & \\
    450 & 4 & 1.26 & 1.02 & 43 & 0.6 & 0.37 & \\
    \midrule
    400 & 9 & 1.17 & 0.98 & 252 & 0.2 & 0.33 & \\
    500 & 9 & & & 293 & 0.8 & 0.34 & 4.241  \\
    600 & 9 & 1.11 & 1.07 & 288 & 1.0 & 0.32 & 4.225  \\
    700 & 9 & & & 274 & 1.2 & 0.27 & 4.211  \\
    800 & 9 & 0.57 & 0.95 & 275 & 1.3 & 0.28 & 4.188  \\
    \bottomrule
    \end{tabular}
    \label{tab:Fit_N2Ar}
\end{table*}

\subsubsection{Characterization}
The optical properties of the prepared samples were characterized ex situ using an M2000X spectral ellipsometer (J.A. Woollam Co.) in the wavelength range from 210  to 1000 nm and an IR-VASE infrared ellipsometer (J.A. Woollam Co.) in the wavenumber range from 250 to 8000 cm\textsuperscript{-1}. The obtained ellipsometric data were analyzed using the commercial software CompleteEase. A three-layer model, substrate-film-roughness, was used to simulate the measurement. The CrN film dispersion functions were parametrized using the number of Gaussian oscillators. The roughness was simulated using the Bruggeman model to approximate the effective medium of a mixture of CrN and void media. The thickness of the “roughness” layer can significantly affect the resulting dielectric function values, due to a strong correlation of these parameters in the fitted model. We included transmittance data in the analysis to decorrelate the model parameters, thus improving the unambiguous determination of the optical function. For this reason, the assumed thickness of the individual CrN films in the gas-composition series was chosen to be sufficiently small to render the layer semi-transparent. The transmittance of CrN samples was measured using an M2000X spectral ellipsometer in transmission mode. Additionally, transmittance was measured over the near-IR range from 500 nm to 1400 nm using a Lambda 1050 spectrophotometer (PerkinElmer). The data obtained were used to refine the absorption constant by point-by-point fitting the aforementioned optical model to the transmittance measurement.

The surface morphology of the CrN coatings was characterized using atomic force microscopy (AFM Dimension ICON, Bruker). Measurements were performed under ambient conditions, and images were obtained using the Peak Force Tapping mode with ScanAsyst air probes (Bruker, radius 2 nm) over areas of $1\times1$ $\mu$m\textsuperscript{2}. 

XRD was performed using a SmartLab SE multipurpose diffractometer (Rigaku Corporation, Tokyo, Japan) equipped with Cu K$\alpha$IQ11 radiation and a HyPix-3000 two-dimensional detector operated in 0D mode. Epitaxial films grown on MgO were analyzed in high resolution by reciprocal space mapping (RSM). The incident optics consisted of a multilayer parabolic mirror, a two-bounce Ge (220) monochromator, and a slit/mask of 0.3 mm and 2 mm, respectively. On the diffracted side, a 0.3 mm slit and a 2.5$\degree$ Soller slit were used. For polycrystalline films deposited on fused silica, the monochromator was replaced with a 2.5$\degree$ Soller slit, and the diffracted arm included a 0.5$\degree$ parallel-slit collimator. Grazing-incidence 2$\theta$ scans were acquired at $\omega$ = 2$\degree$.  A beam plate was mounted on the diffracted arm to suppress parasitic surface scattering and improve the signal-to-noise ratio. Phase analysis was carried out in SmartLab Studio II using the PDF-5 database and Rietveld refinements by TOPAS V3.

Resistivity measurements were carried out using the differential van der Pauw (vdP) method in a quasi-square arrangement at room temperature (298 K), using a Keithley 6221 current source and two electrometers, a Keithley 6514 with a Keithley 2182A nanovoltmeter, which recorded the voltage difference between the electrometers, together with a Keithley 708 B switching matrix. The linearity of the contacts was checked for all measured samples to ensure an ohmic contact. The Hall effect was measured using the same instrument in a magnetic field of +0.8 T and -0.8 T.

\subsection{Theoretical}
\label{subsec:DFT}
Theoretical studies are performed for 2D bilayer CrN(001), CrN$_{1-x}$(001), and Cr$_{1-x}$N(001) with $x=3$ at\%. The compositions are modeled by supercells created from the bulk rocksalt-type CrN structure with an experimental lattice parameter of 4.15\AA. A vacuum space of 10~\AA \ is added perpendicularly to the top and bottom of the bilayer to avoid interactions between the bilayer and its periodic images \cite{xie,fu}. The bilayer CrN(001) is composed of 128 atoms (64 Cr and 64 N). Single, neutral Cr and N vacancies are introduced into each monolayer to produce a vacancy concentration of 3 at~\%. An antiferromagnetic order is imposed on the Cr sublattice in the bilayer, with ferromagnetically coupled Cr atoms within each monosheet and antiferromagnetically coupled sheets \cite{zhang}.  

The electronic and dielectric properties of the 2D bilayer CrN(001), CrN$_{1-x}$(001), and Cr$_{1-x}$N(001) are investigated within the spin-polarized density functional theory (DFT) implemented in the \textsc{VASP} code \cite{vasp1,vasp2}. Calculations employ: (i) the generalized gradient approximation (GGA) with parameterization of Perdew, Burke, and Ernzerhof for the exchange correlation potential \cite{pbe}, (ii) the projector-augmented wave (PAW) method to describe electron-ion interaction \cite{paw1,paw2}, (iii) a plane-wave basis set with cutoff energy of 520 eV along with the PAW pseudopotentials with the following configurations of valence electrons: Cr($3d^5 4s^1$), N($2s^2 2p^3$). The localized character of Co-$3d$ electrons is taken into account via the DFT+$U$ formalism and the Dudarev \emph{et al.} \cite{dudarev} approach. The strong on-site Coulomb interaction in the Co $3d$ shell is described by the effective Hubbard potential $U=4.5$~eV, determined self-consistently for the CrN(001) bilayer using the approach of Timrov \emph{et al.} based on the density functional perturbation theory (DFPT) implemented in the \textsc{quantum espresso} package (QE) \cite{qe1,qe2,qe3}.  We note that the same value of the effective Hubbard potential has recently been used in theoretical studies of electronic correlations in epitaxial CrN thin films \cite{kalal}. The atomic positions in the modeled structures are optimized, and their Brillouin zones are sampled with the $6\times6\times1$ $k$-point mesh. The semi-empirical corrections of Grimme et al. \cite{grimme} are incorporated to account for van der Waals interactions, which are important in layered structures \cite{vidal}.

\section{Results and Discussion}
\subsection{Structural analysis}
\begin{figure*}
    \centering
    \begin{subfigure}[b]{0.32\textwidth}
        \centering
        \includegraphics[width=\textwidth]{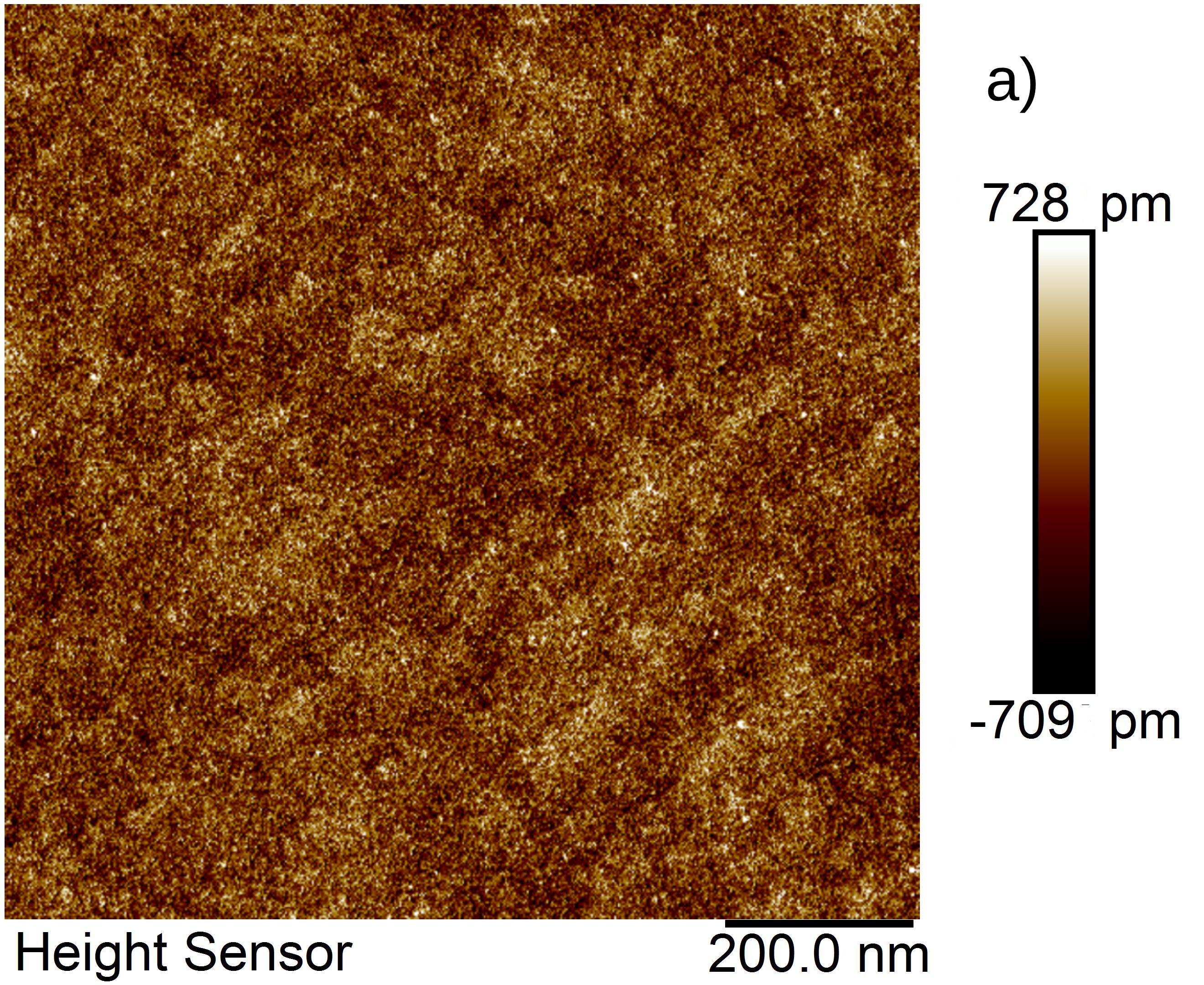}
    \end{subfigure}%
    \begin{subfigure}[b]{0.32\textwidth}
        \centering
        \includegraphics[width=\textwidth]{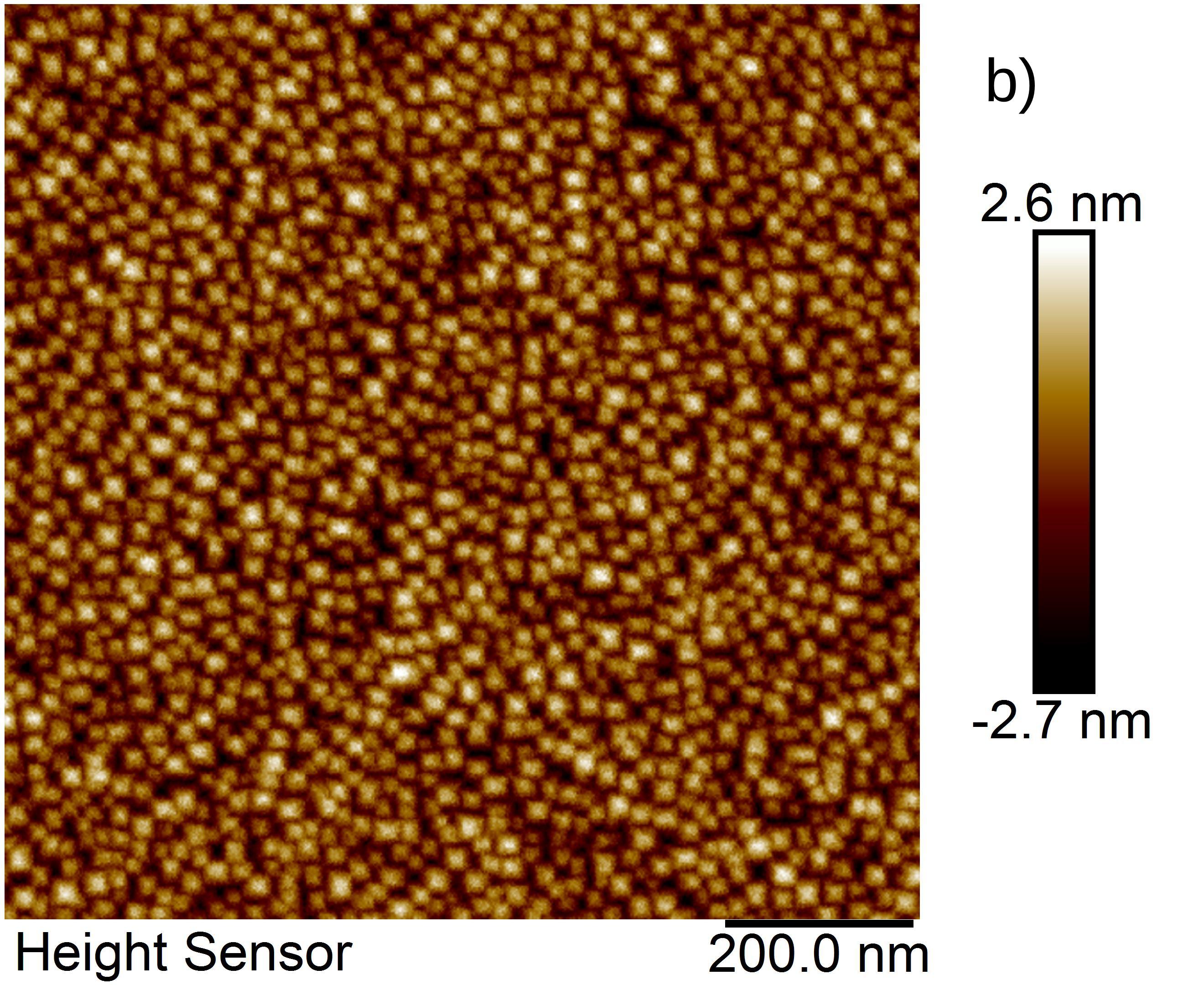}
    \end{subfigure}%
    \begin{subfigure}[b]{0.32\textwidth}
        \centering
        \includegraphics[width=\textwidth]{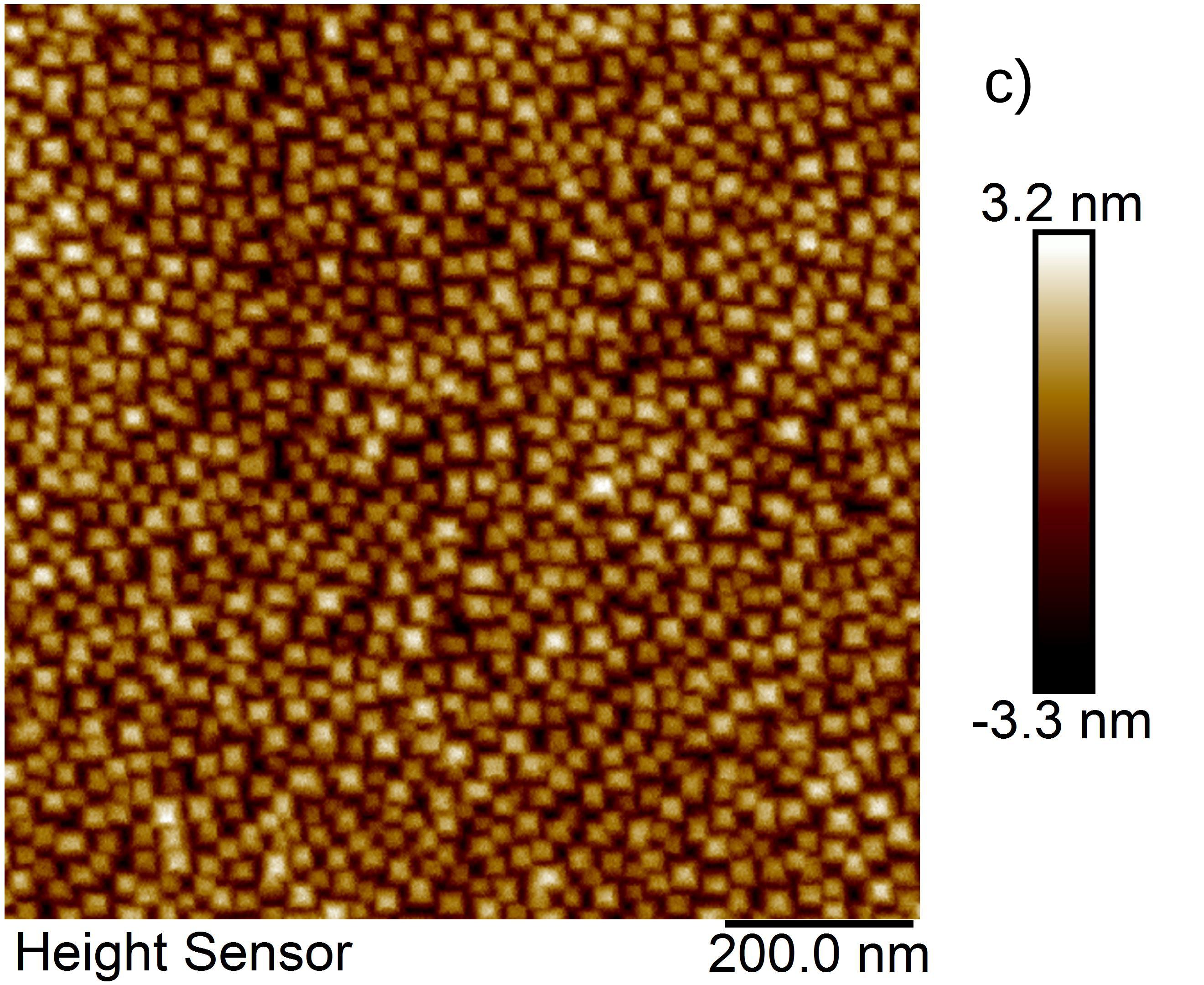}
    \end{subfigure}%
    
    \begin{subfigure}[b]{0.32\textwidth}
        \centering
        \includegraphics[width=\textwidth]{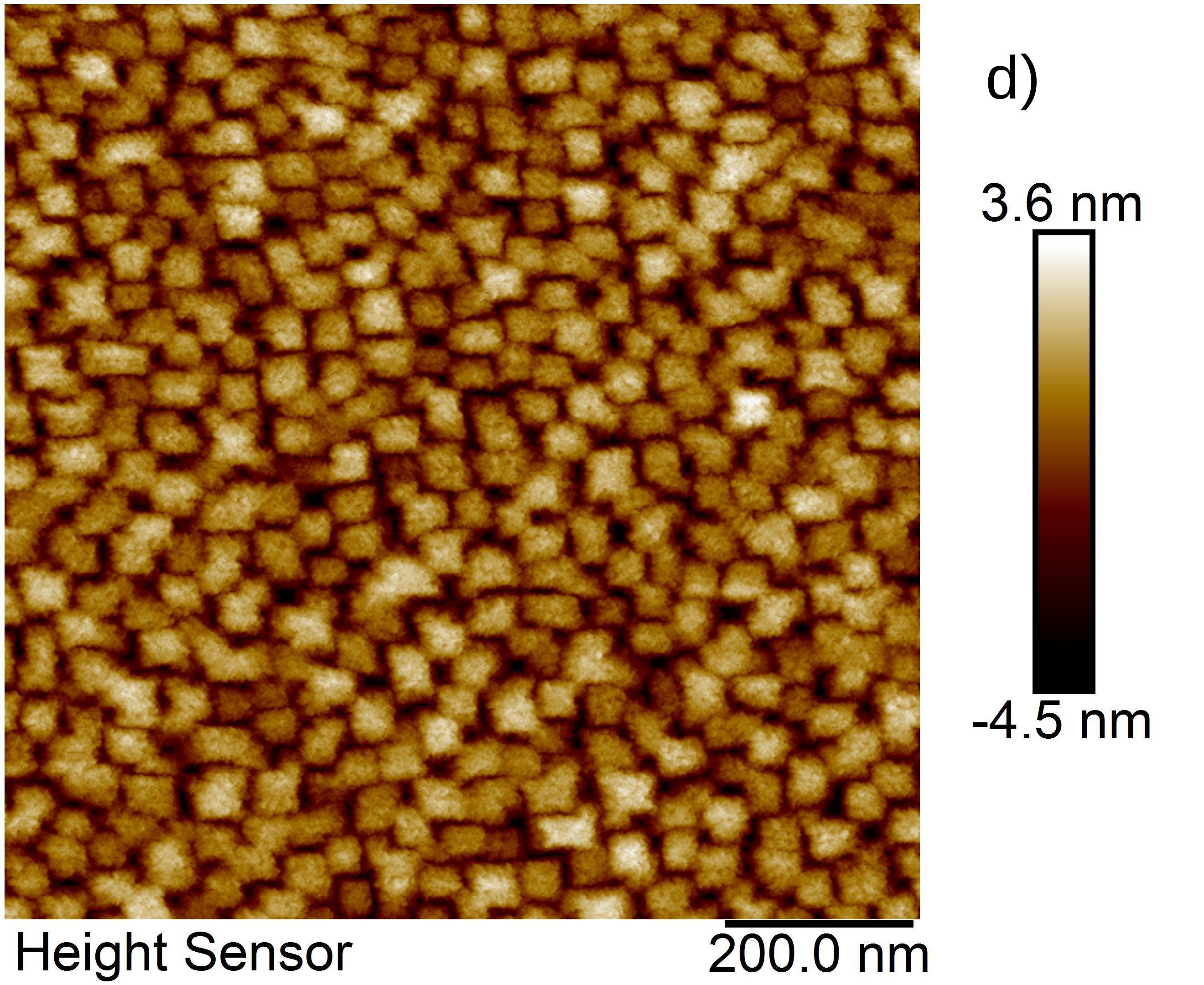}
    \end{subfigure}%
    \begin{subfigure}[b]{0.32\textwidth}
        \centering
        \includegraphics[width=\textwidth]{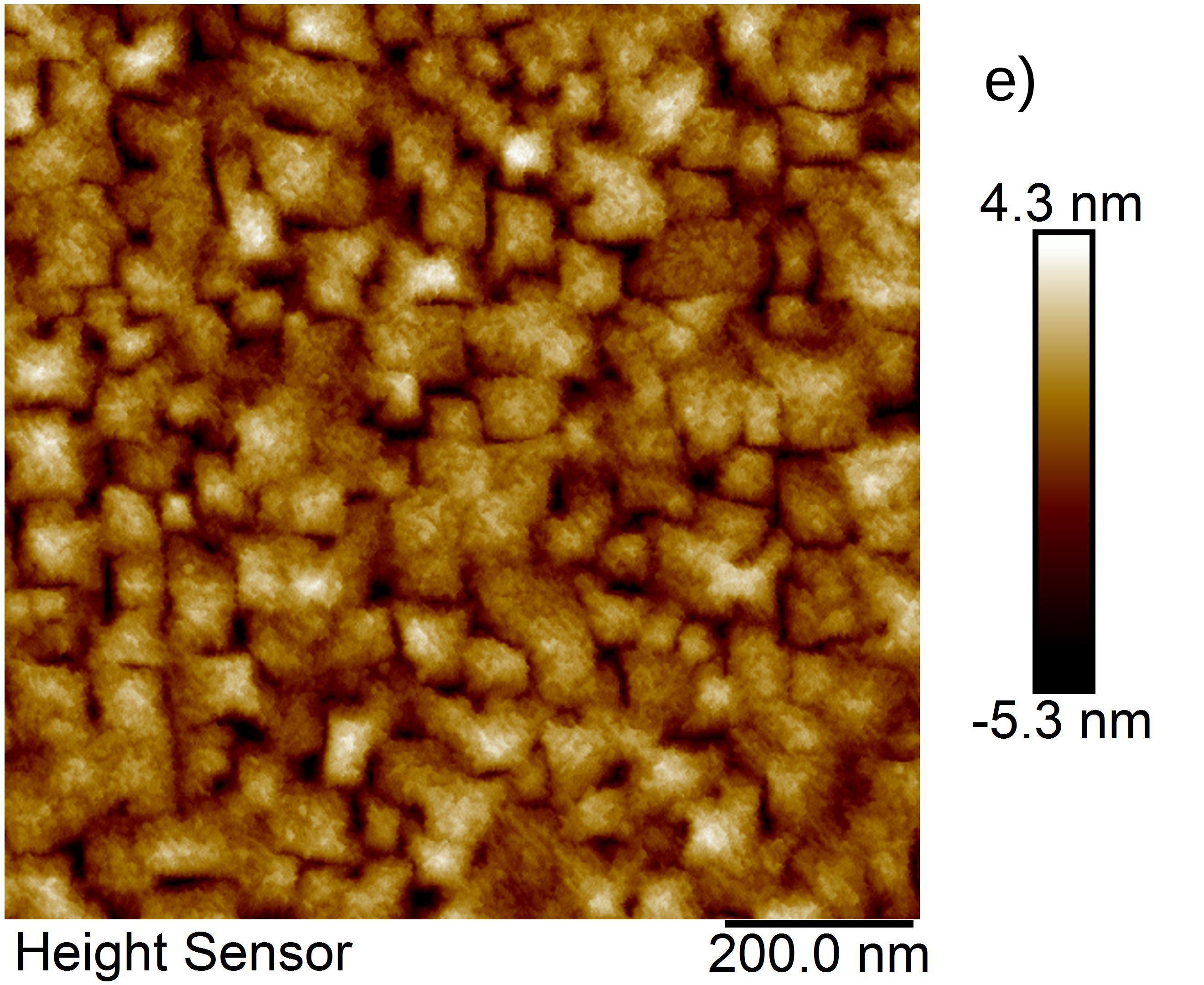}
    \end{subfigure}
    \begin{subfigure}[b]{0.32\textwidth}
        \centering
        \includegraphics[width=\textwidth]{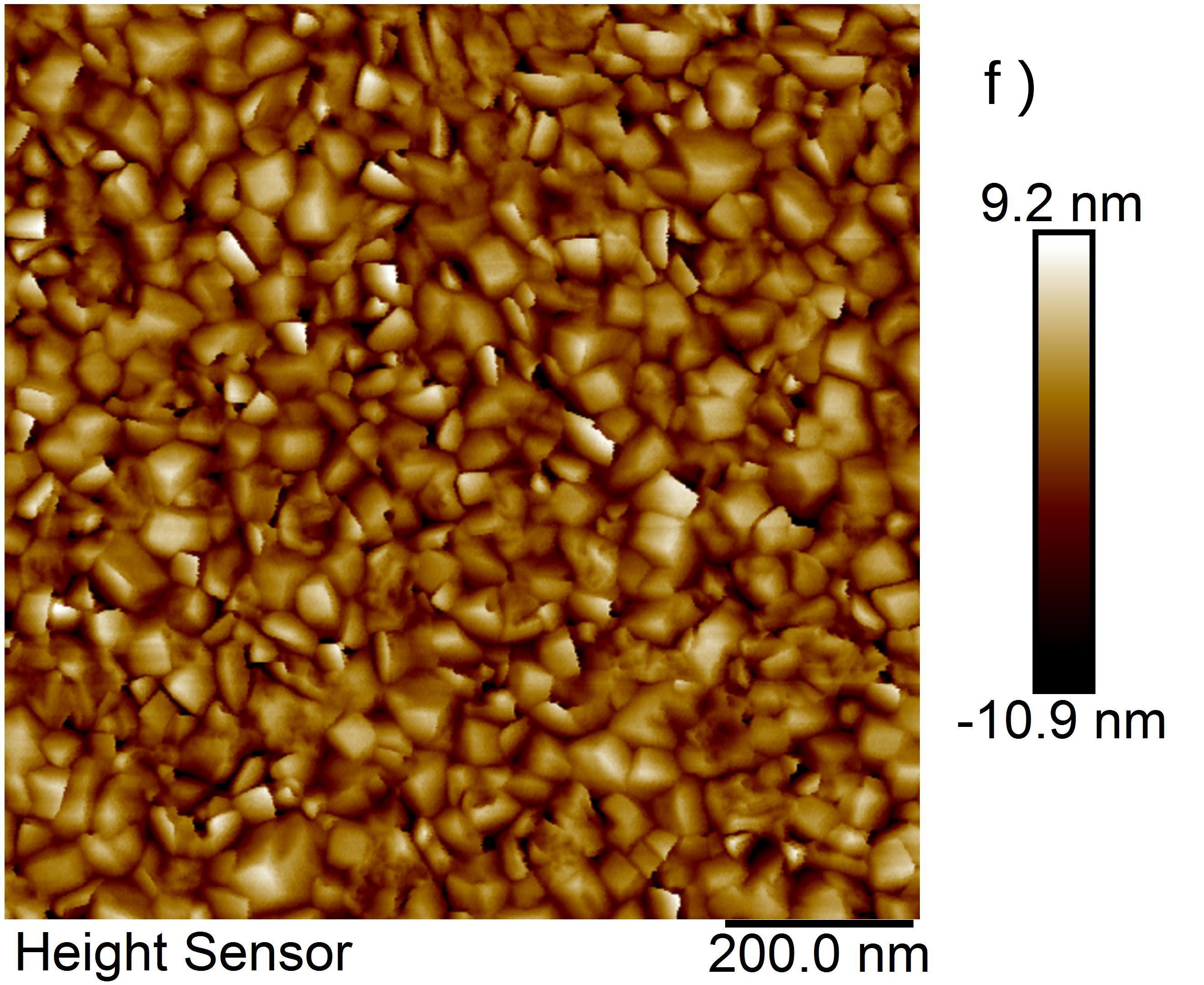}
    \end{subfigure}%
    \caption{AFM images of the surface morphology of CrN layers deposited on MgO substrate at temperatures of (a) 400$\degree$C, (b) 500$\degree$C, (c) 600$\degree$C, (d) 700$\degree$C, and (e) 800$\degree$C. (f) AFM image of CrN layer deposited on fused silica substrate at 700 $\degree$C}.
    \label{fig:AFM2}
\end{figure*}

The structure of a material has a significant impact on its resulting physical properties. This structure can be influenced by the substrate material and its temperature during layer growth. This effect is well indicated by the surface morphology as demonstrated in the AFM images in Fig. \ref{fig:AFM2}. The images marked (a) to (e) represent samples deposited at constant temperature from 400 to $800\degree$C on an MgO substrate. The $N_2/Ar$ gas flow ratio was set to 4 in this case. The increase in grain size is clearly evident as substrate temperature increases. For comparison, Fig.\ref{fig:AFM2} (f) shows a CrN film deposited on a fused silica substrate at $700\degree$C. The morphology of this CrN film exhibits randomly arranged grains, while the surface of the CrN layer on the MgO substrate shows a certain degree of oriented arrangement. Although the rectangular grains are aligned along the MgO structural axis, a slight degree of misorientation is evident. The CrN layer deposited at $400\degree$C on the MgO substrate has a very fine grain structure, leading to a very smooth surface (Fig. \ref{fig:AFM2} a). With increasing temperature, the square grain size gradually increases, reaching 20 nm, 25 nm, 40 nm, and 70 nm for the $500\degree$C, $600\degree$C, $700\degree$C, and $800\degree$C samples, respectively. The corresponding RMS roughness is shown in Table \ref{tab:Fit_N2Ar}.

\begin{figure}
    \centering
    \includegraphics[width=1.0\linewidth]{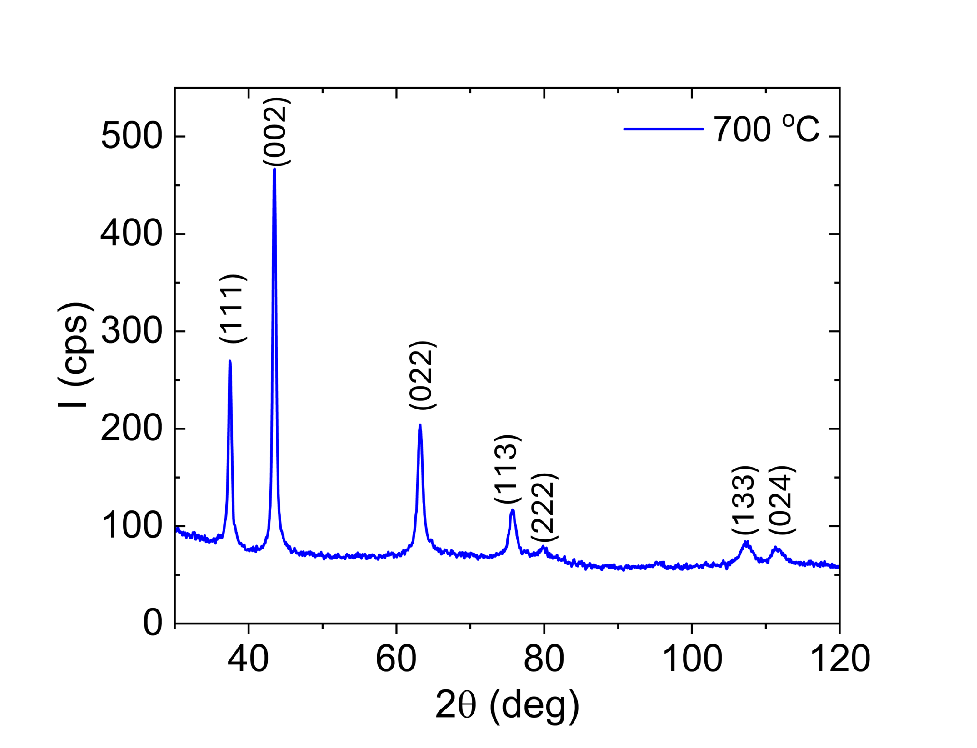}
    \caption{XRD pattern of CrN deposited on fused silica (SL) at 700 $\degree$C. The peaks are indexed according to their (\textit{hkl}) Miller indices.}
    \label{fig:CrN_FS-XRD}
\end{figure}

XRD analysis can provide deeper insight into the material's structure. Polycrystalline samples showed similar results regardless of the deposition temperature. Figure \ref{fig:CrN_FS-XRD} presents the data for the film deposited at $700\degree$C  as a representative example. The only detected phase corresponds to cubic CrN ($Fm\bar{3}m$) with a refined lattice parameter of 4.166 \si{\angstrom}.

\begin{figure*}
    \centering
    \includegraphics[width=0.8\linewidth]{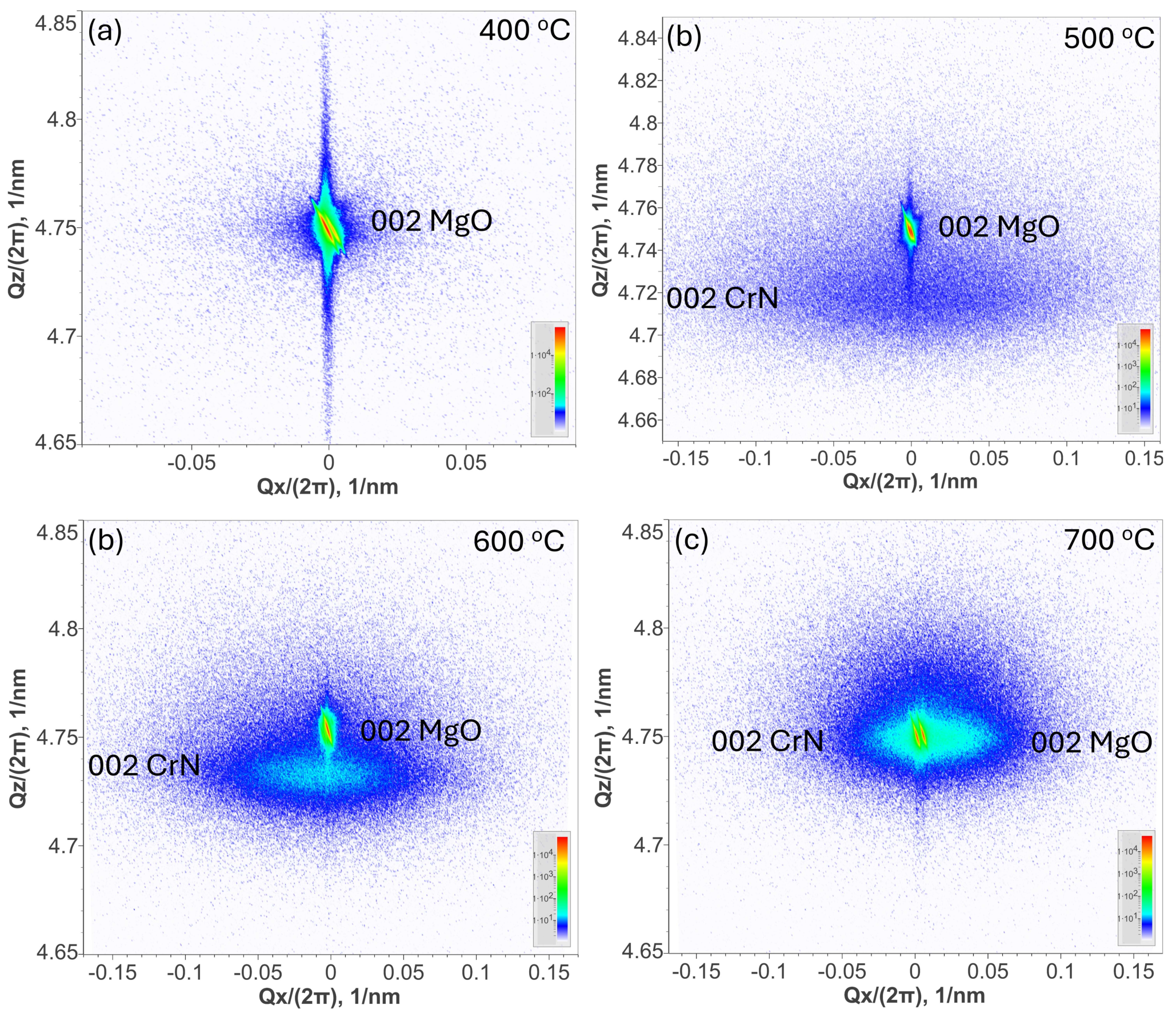}
    \caption{ Symmetric 002 reciprocal space maps of CrN layers deposited on MgO at temperatures of (a) $400\degree$C, (b) $500\degree$C, (c) $600\degree$C, (d) $700\degree$C}
    \label{fig:RSM_CrN-Temp}
\end{figure*}

\begin{figure*}
    \centering
    \includegraphics[width=0.8\linewidth]{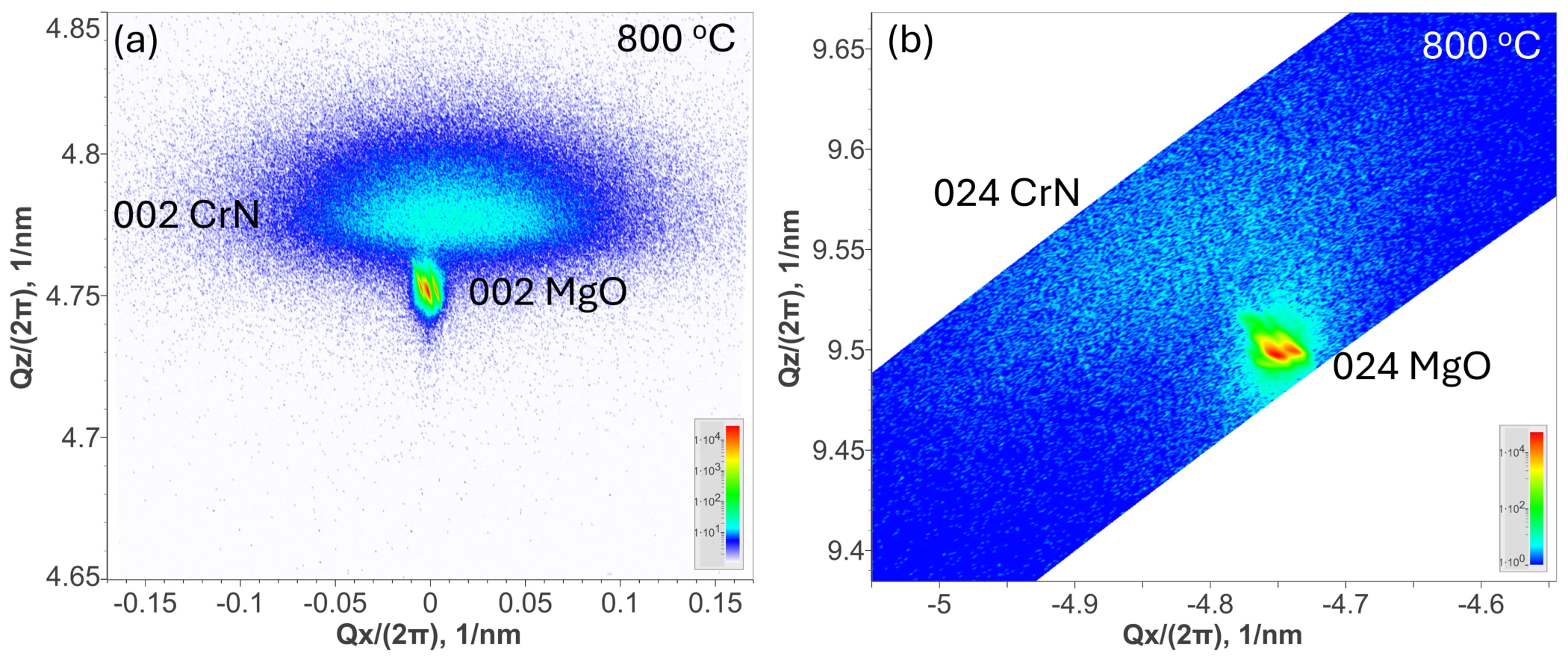}
    \caption{Reciprocal space maps of the CrN layer deposited on MgO at $800\degree$C: (a) symmetric 002 and (b) asymmetric 024 reflections}
    \label{fig:RSM_CrN-800C}
\end{figure*}

For films grown on MgO, a clear contraction of the CrN lattice is observed as the deposition temperature increases. The out-of-plane lattice parameter was obtained from symmetric reciprocal space maps (see Fig.\ref{fig:RSM_CrN-Temp} and Fig.\ref{fig:RSM_CrN-800C}(a)). Due to the very low diffracted intensity of the films, asymmetric reciprocal space mapping was only possible for the sample grown at 800$\degree$C; see Fig.\ref{fig:RSM_CrN-800C}(b). The results of the analysis shown in Table \ref{tab:Fit_N2Ar} indicate a decrease in the lattice parameter with increasing substrate temperature during deposition. Although the in-plane lattice parameter could not be reliably extracted from the asymmetric map due to the weak film signal, the film and substrate peaks are observed at the same azimuth. This result, together with the square-shaped grains visible in the AFM images, indicates that the film grows epitaxially on the substrate. This phenomenon is referred to as cube-on-cube epitaxial growth \cite{gall_growth_2002}. 

The chemical composition of CrN layers was analyzed using XPS. N 1s and Cr 2p core-level spectra were recorded to determine the N/Cr stoichiometric ratio. 
Owing to the high surface sensitivity of XPS, the as-deposited CrN films exhibit significant surface contamination, with approximately 20 at.\% carbon and 15–30 at.\% oxygen. Consequently, the measured N/Cr ratio may be influenced by surface contamination and oxidation. Most of the contamination is removed after Ar$^+$ sputtering, thereby allowing analysis of the composition of the underlying CrN layer. The compositional analysis of both the as-deposited and sputtered films is summarized in Table \ref{tab:Fit_N2Ar}. In both cases, the N/Cr ratio increases with increasing N$_2$/Ar gas flow ratio during deposition. However, the sputtered samples exhibit systematically lower N/Cr ratios than the as-deposited films. This difference is most likely due to preferential sputtering of nitrogen during Ar$^+$ ion etching, leading to an underestimation of the nitrogen concentration. The temperature series shows an N/Cr ratio of approximately 1.1 at the untreated surface, except for the CrN layer deposited at 800 $\degree$C, which exhibits a significant decrease in nitrogen content within the CrN structure. A certain loss of nitrogen is evident in CrN deposited above 700 $\degree$C, as reported in publication \cite{gall_growth_2002}.

\begin{figure}
    \centering
    \includegraphics[width=1.0\linewidth]{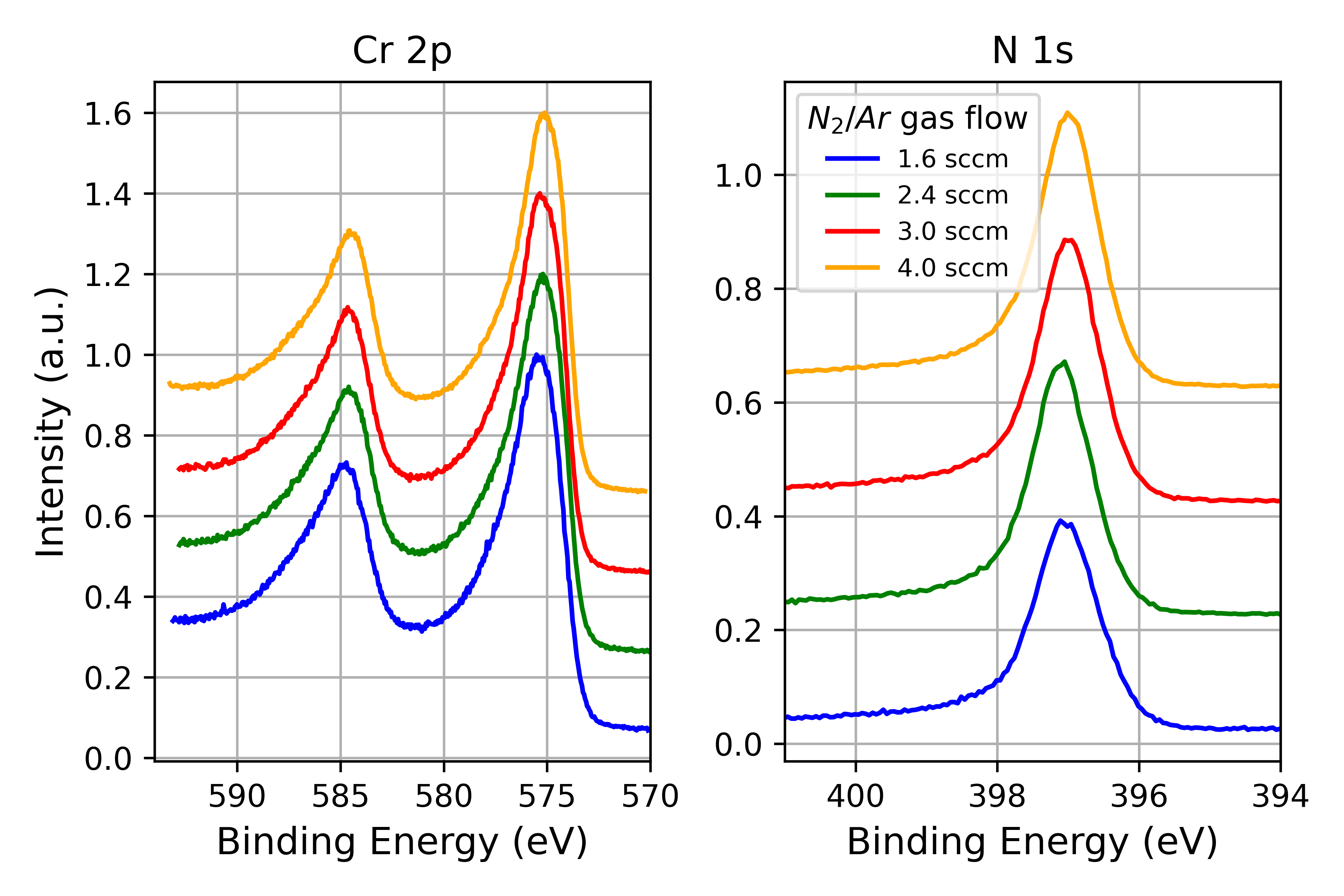}
    \caption{XPS core levels Cr 2p and N 1s after 1 keV Ar+ sputtering}
    \label{fig:XPS_N2Ar-flow_sputtered}
\end{figure}

 The Cr 2p and N 1s spectra are shown in Fig. \ref{fig:XPS_N2Ar-flow_sputtered}. Their exact positions were corrected with respect to the Fermi-edge position in the valence-band measurement. Nevertheless, the exact correction is difficult to determine due to the overlap between the Fermi level and the Cr 3d states. The C 1s and O 1s bands were also present in the measured spectra. However, after sputtering with Ar ions, the intensity of these bands decreased significantly, indicating that these elements are present mainly on the sample surface.  

\subsection{Electron-transport properties}
\begin{figure}
    \centering
    \includegraphics[width=1.0\linewidth]{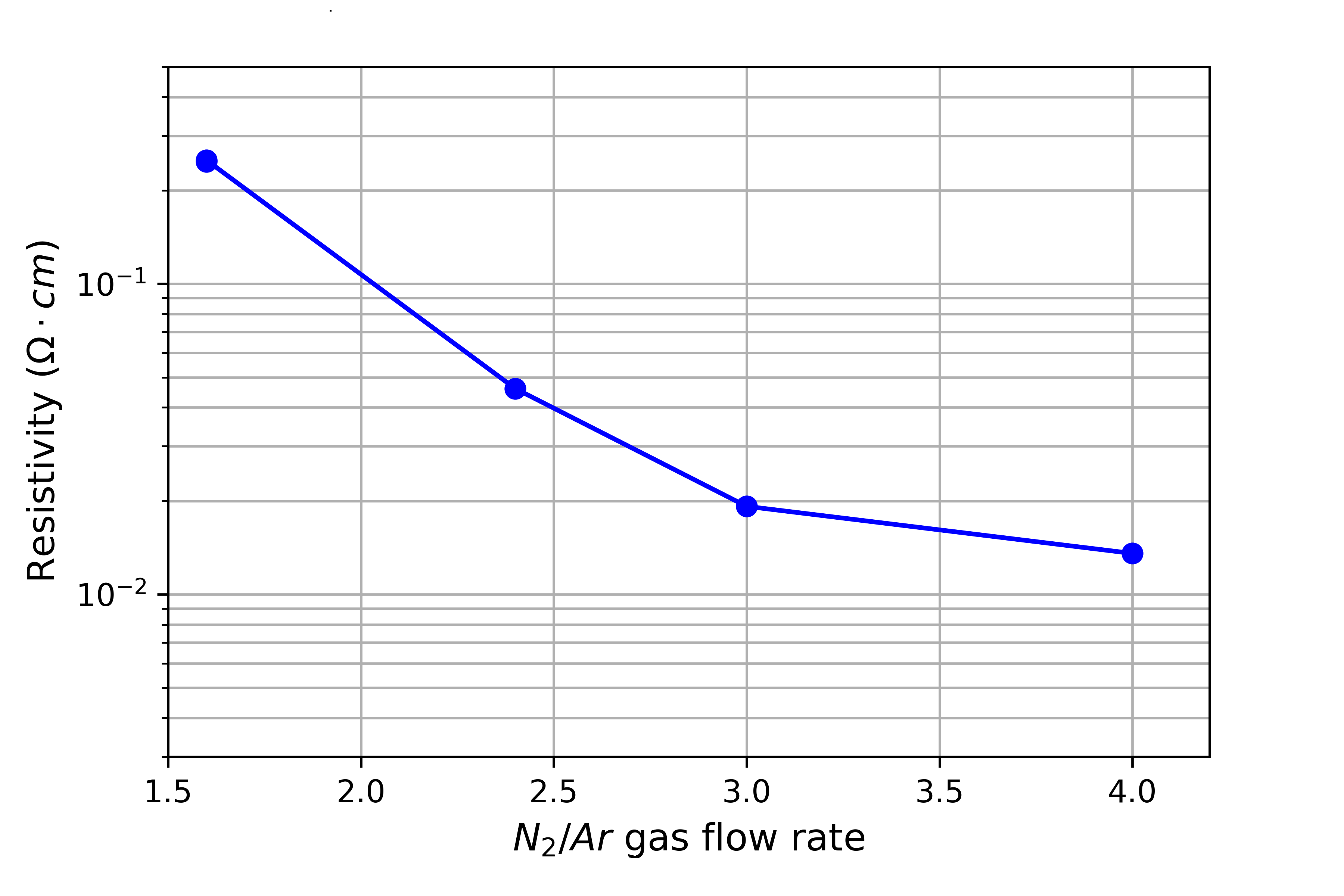}
    \caption{Electrical resistivity of CrN films deposited on a fused silica substrate versus N$_2$/Ar gas flow ratio used during the deposition. Substrate temperature during deposition was 450°C}
    \label{fig:elres_N2Ar}
\end{figure}

Transition-metal nitrides are interstitial compounds. Therefore, they tend to be nonstoichiometric. A small deviation in composition can mainly affect their optical and electron transport properties. Therefore, we first optimized the stoichiometry of the CrN structure with respect to optical and electron transport properties. Stoichiometry was controlled by the nitrogen-to-argon gas ratio in the working gas, as shown in Table \ref{tab:Fit_N2Ar}. The evolution of electrical resistivity is illustrated in Fig. \ref{fig:elres_N2Ar}.  The electrical resistivity decreases with increasing nitrogen content in the structure, as seen in the graph, where the resistivity decreases from  $2.5\times 10^{-1}$ to $1.4 \times 10^{-2}$ $\Omega \cdot$cm when the gas ratio N$_2$/Ar during deposition increases from 1.6 to 4. The following section focuses on super-stoichiometric layers with a high nitrogen content.

\begin{figure}
    \centering
    \includegraphics[width=1.0\linewidth]{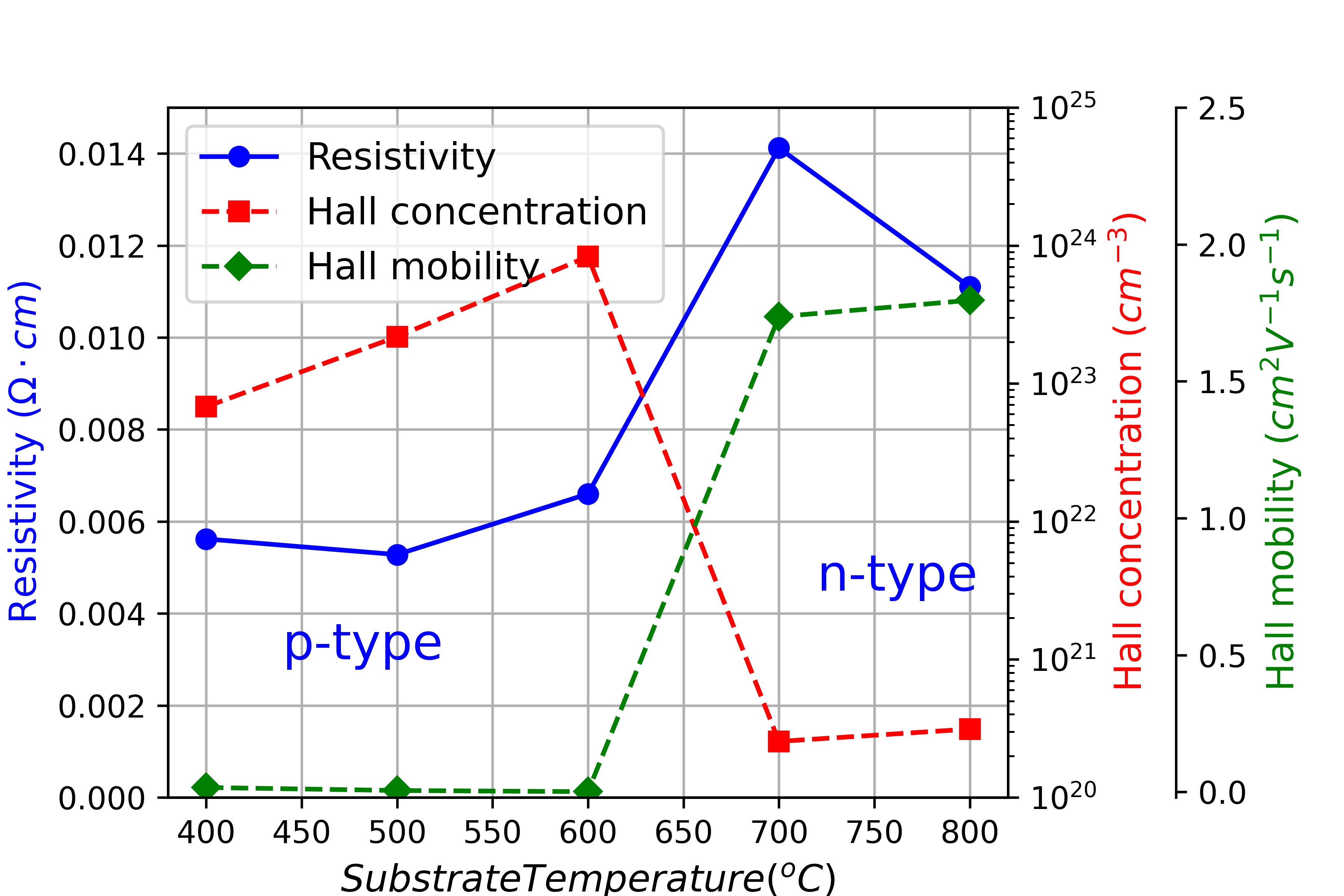}
    \caption{Dependence of electrical resistivity, Hall concentration, and Hall mobility of CrN films on substrate temperature used during the deposition process.}
    \label{fig:elres_Ts}
\end{figure}

The substrate temperature is an important parameter for the crystal growth of CrN. We have carried out CrN deposition in a $N_2$/Ar gas mixture (9) at substrate temperatures from $400\degree$C to $800\degree$C, which influenced changes in crystallinity, as discussed in the following subsection. At the same time, the substrate temperature also affects the electrical conductivity and its mechanism. The electrical resistivity below $600\degree$C is about $5.5 \times 10^{-3} \space  \Omega \cdot$ cm (see Fig. \ref{fig:elres_Ts}). The electrical conductivity is assumed to be the product of the concentration, the mobility, and the charge of the free charge carriers. A p-type conductivity mechanism was detected for these samples by Hall-effect measurement. Hence, the mechanism of electrical conductivity is mediated through free holes. The hole concentration increases with substrate temperature below $600\degree$C, while their mobility is relatively low (see Fig. \ref{fig:elres_Ts}). Above $600\degree$C, the concentration of free charge carriers decreases significantly. This may be related to the decrease in structural defects. At the same time, the scattering of free charge carriers in the defect-free structure decreases, resulting in significantly increased mobility. The increased mobility is offset by a significant decrease in the concentration of charge carriers, ultimately leading to an increase in electrical resistivity (see Fig. \ref{fig:elres_Ts}).
At the same time, the conduction mechanism changes to n-type, mediated by free electrons. This change likely also contributes to the increase in mobility because holes have a higher effective mass than electrons. The result is a transition to electronic conduction, characterized by enhanced charge-carrier mobility. In \cite{gall_growth_2002}, the authors report on understoichiometric films with reduced nitrogen concentration caused by nitrogen desorption during deposition at elevated temperatures above $700 \degree$C. This desorption reduces the number of defects in the structure, thereby decreasing the concentration of charge carriers, consistent with our results.  
 
\subsection{Optical properties}
\begin{figure}
    \centering
    \includegraphics[width=1.0\linewidth]{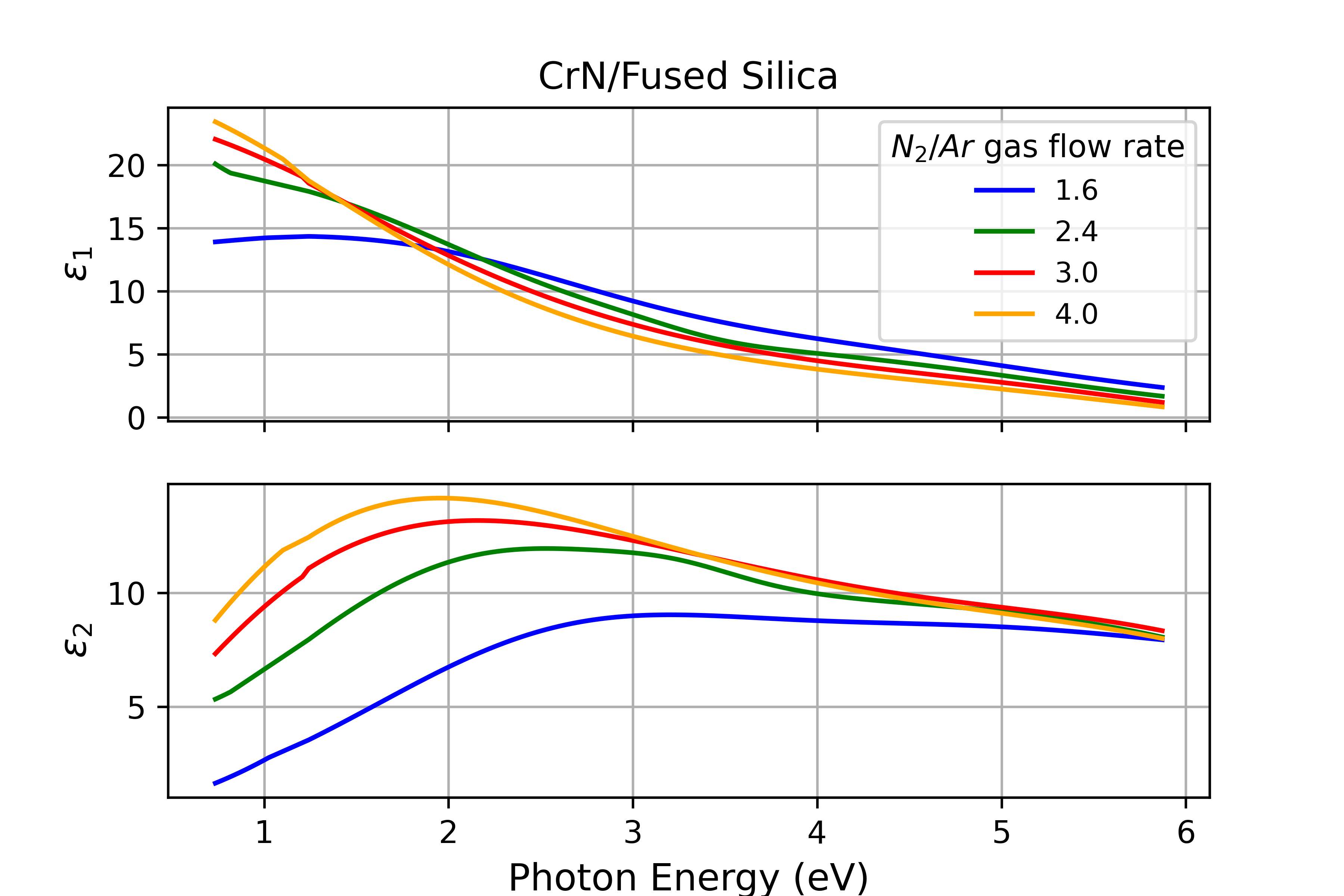}
    \caption{Dielectric functions of CrN film on fused silica substrate. The films were deposited at various N$_2$/Ar gas flow ratios at a substrate temperature of $450\degree$C. The thickness of the CrN film was estimated between 33 and 43 nm.}
    \label{fig:eps12_N2Ar}
\end{figure}

The optical properties of the chromium nitride (CrN) films were investigated through the analysis of the dielectric function, with both real and imaginary parts examined from the ultraviolet (UV) to near-infrared (NIR) spectral region, as shown in Fig. \ref{fig:eps12_N2Ar}. The graph is the result of merging measurements from the UV-Vis-NIR and IR ellipsometers. The dispersion model of the CrN layer was described using a general oscillator model. We used three Gaussian oscillators, assuming a certain statistical broadening. The resulting parameters of the dispersion function are shown in Table \ref{tab:Genosc_N2Ar}. The obtained film thickness values are shown in Table \ref{tab:Fit_N2Ar}. The main difference among the curves is in the low-energy part below 3 eV. This variation is mainly reflected in the amplitude $A_1$ and the energy position $E_1$ of the first oscillator: for CrN deposited at a higher N$_2$/Ar gas flow ratio, the amplitude increases, while the oscillator position shifts to lower energies.

\begin{table}
\small
\centering
\caption{Results of analysis of ellipsometric data of CrN films deposited at various N$_2$/Ar gas flow ratios. The $A_i$, $\Gamma_i$, and $E_i$ with $i=1,2,3$ represent the amplitude, broadening, and energy position of the three Gaussian oscillators used in the dispersion model.}
\resizebox{0.48\textwidth}{!}{
\begin{tabular}{cccccccccc}
    \toprule
    N$_2$/Ar & $A_1$ & $\Gamma_1$ & $E_1$ & $A_2$ & $\Gamma_2$ & $E_2$ & $A_3$ & $\Gamma_3$ & $E_3$ \\
    \midrule
    1.6 & 4.5 & 2.3 & 2.5 & 3.5 & 3.3 & 4.6 & 17.4 & 30.5 & 5.7 \\
    2.4 & 7.0 & 2.5 & 2.0 & 0.5 & 0.8 & 3.3 & 9.1 & 5.5 & 4.7 \\
    3 & 13.7 & 3.3 & 0.6 & 8.7 & 3.6 & 3.1 & 6.7 & 3.5 & 6.0 \\
    4 & 13.3 & 3.2 & 0.7 & 9.2 & 4.2 & 3.1 & 5.3 & 3.4 & 6.2 \\
    \bottomrule
    \end{tabular}
    }
    \label{tab:Genosc_N2Ar}
\end{table}

The real part of the dielectric function, $\varepsilon _1$, which represents the electronic polarizability of the material, showed a steady decrease from a value of 13.9 - 23.4 at a photon energy of 0.73 eV (corresponding to a wavelength of 1690 nm) to a value ranging between 0.8 and 2.4 at a photon energy of 5.9 eV (210 nm). This trend suggests that the CrN film exhibits a significant refractive response in the NIR region. The real and imaginary parts of the dielectric function are interconnected by the Kramers-Kronig relations. The imaginary part of the dielectric function, $\varepsilon_2$, which is associated with absorption-induced optical losses, also shows a notable response across this spectral range. It rises at lower energies (starting at 0.73 eV) and peaks at photon energies between 1.9 and 3.5 eV, depending on the N$_2$/Ar gas flow rate. Above this energy, the curves gradually decrease to about 8 at 5.9 eV. This feature suggests a semi-transparent region in the near-infrared below 1 eV. A very similar course of optical functions of CrN as in our sample for $N_2$/Ar=4 is shown in publication \cite{aouadi_spectroscopic_2001} for pure CrN. 

\begin{figure}
    \centering
    \includegraphics[width=1.0\linewidth]{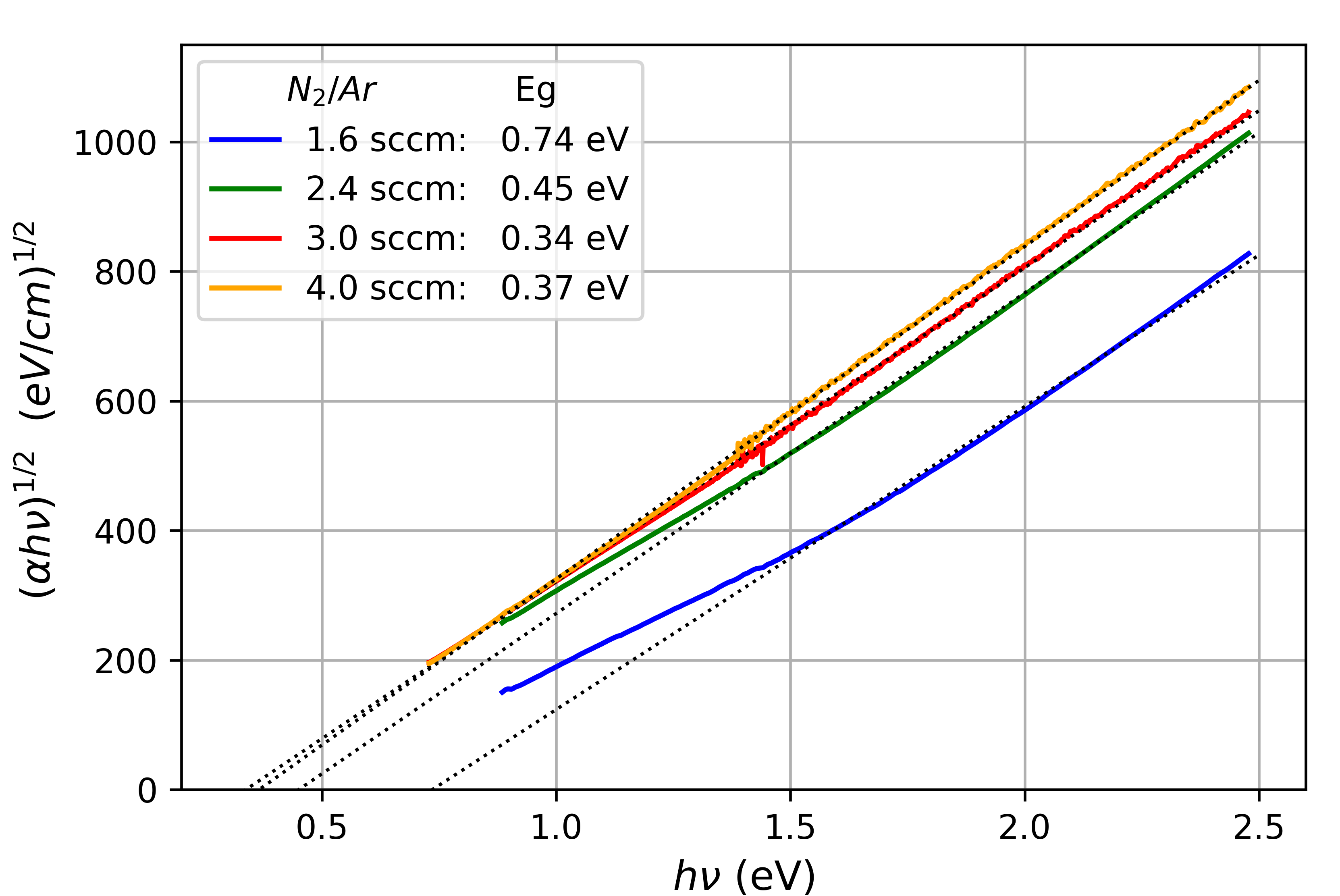}
    \caption{Tauc plot of CrN films deposited at various N$_2$/Ar gas flow ratios. The estimated band-gap energy is depicted in the graph, considering the extrapolation error as large as $\pm$0.05 eV.}
    \label{fig:Taucplot}
\end{figure}

The initial increase in the value of $\varepsilon _2$ corresponds to an absorption edge indicating the band gap energy of CrN. There is a shift in the absorption edge among the samples, associated with changes in the band gap energy. The absorption coefficient ($\alpha$) obtained from CrN samples was used to determine the band-gap energy by Tauc-plot extrapolation. The absorption coefficient is related to the band gap energy (E$_g$) using the linear equation $(\alpha h \nu )^{\beta}=\gamma(h\nu-E_g)$, where $\beta$ possesses values of 2 and 1/2 for direct and indirect transitions across the band gap, respectively. The Tauc plot of CrN films is shown in Fig. \ref{fig:Taucplot}, considering CrN as an indirect band gap semiconductor. The band-gap energy decreases from about 0.74 to 0.34 eV with increasing nitrogen content in the CrN structure (see Table \ref{tab:Fit_N2Ar}). The upper value is comparable to the 0.75-0.8 eV reported in \cite{alam_electronic_2022}. The behavior of optical functions in the near-infrared region, represented by the band-gap energy, can be correlated with changes in electrical resistivity shown in Fig. \ref{fig:elres_N2Ar}. CrN, with the largest band gap, has the highest electrical resistance. As the electrical resistance decreases due to increasing defect concentration, the band-gap energy decreases. The increased concentration of defects in the structure creates acceptor states in the band gap. These states can absorb photons in the sub-band-gap region and affect the absorption spectrum, which can be interpreted as a decrease in the band gap energy.

\begin{figure}
    \centering
    \includegraphics[width=1.0\linewidth]{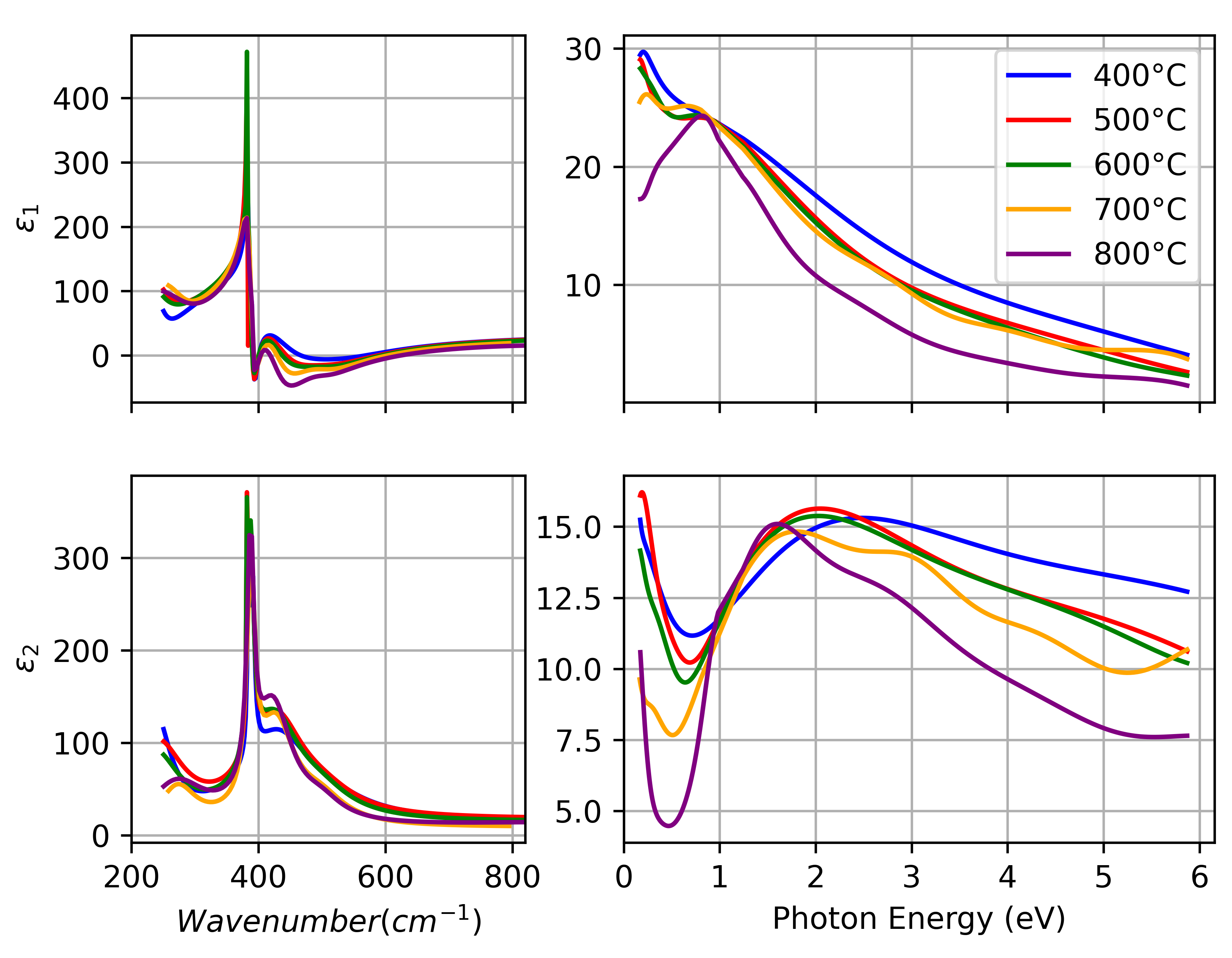}
    \caption{Dielectric functions of CrN film on fused silica substrate. The films were deposited at substrate temperatures from $400\degree$C to 800$\degree$C. The N$_2$/Ar gas flow ratio was 4. The thickness of the CrN film was estimated between 250 and 290 nm.}
    \label{fig:eps12-T}
\end{figure}

The dependence of the dielectric function on photon energy becomes even more apparent in the temperature series of thicker CrN layers with a thickness of approximately 400 nm (see Fig. \ref{fig:eps12-T}). These films were deposited at substrate temperatures ranging from 400 $\degree$C to 800 $\degree$C. Due to their sufficient thickness, the films exhibit interference oscillations in the semitransparent window in the near-infrared region. Therefore, ellipsometric data acquired by UV-Vis and IR ellipsometry were fitted simultaneously to enable the precise determination of the film thickness, which is shown in Table \ref{tab:Fit_N2Ar}. In the visible spectral range, the dielectric response exhibits the same photon energy dependence as described above. The most pronounced differences among the samples deposited at various temperatures are observed in the near-infrared region, where CrN exhibits semitransparent behavior. This effect is particularly evident in the imaginary part of the dielectric function ($\varepsilon_2$) below the optical band gap, where a significant decrease in $ \varepsilon_2$ is observed. The magnitude of this decrease strongly depends on the substrate temperature during deposition. The film grown at $400\degree$C shows only a slight reduction in $\varepsilon_2$, whereas the effect becomes progressively more pronounced as the temperature increases, reaching its maximum for the film deposited at $800\degree$C. 
This behavior indicates the presence of localized electronic states within the band gap, which are associated with structural disorder. Such states may arise from defects in the crystal lattice, whose concentration decreases as the substrate temperature increases. This phenomenon also reflects the increasing electrical resistivity for the temperature range between 500 and 700 $\degree$ C, as shown in Fig. \ref{fig:elres_Ts}.

With further decrease in photon energy, the imaginary component of the dielectric function increases significantly as vibrational processes take place, as shown in the left part of Fig. \ref{fig:eps12-T}. In the far-infrared (FIR) spectral region, the dielectric function of CrN exhibits a distinct vibrational character, highlighted by pronounced absorption features. A strong vibrational band appears around 400 cm$^{-1}$ in the imaginary part of the dielectric function, indicating the presence of active phonon modes within this range. The band consists of two peaks located at 385 and 430 cm$^{-1}$. A similar vibrational absorption band at a photon energy of 48.7 meV (corresponding to 393 cm$^{-1}$) is reported in Ref.~\cite{zhang_crn_2010}.

Taking into account the Kramers-Kronig relations, the associated absorption peak results in a reduction of the real part of the dielectric function, which becomes even negative between approximately 450 and 600 cm$^{-1}$. This behavior reflects a metallic-like optical response arising from the strong coupling of infrared radiation with lattice vibrations. The magnitude of this effect increases systematically with the deposition temperature, reaching its maximum in the film grown at $800\degree$C.

\subsection{Ab initio modeling}
\begin{figure}
    \centering
    \includegraphics[width=0.5\textwidth]{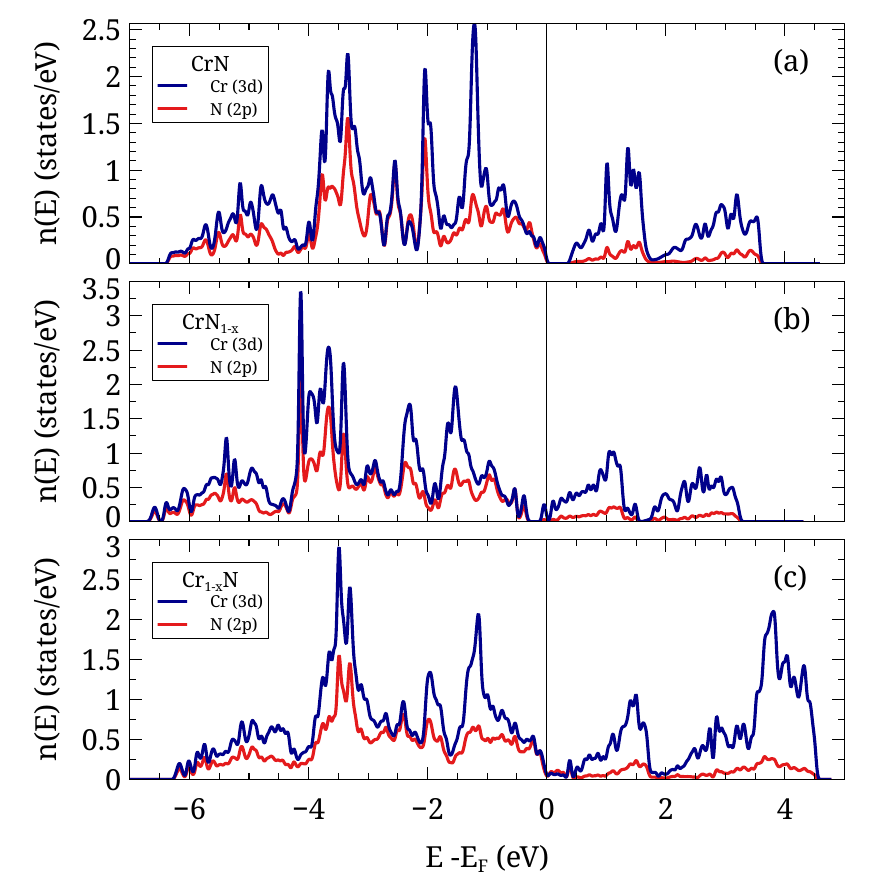}
    \caption{Atomic and orbital projected density of electronic states for (a) CrN, (b) CrN$_{1-x}$, and (c) Cr$_{1-x}$N bilayers. The density of states is shown for one monolayer, as the density of states in a monolayer with reversed direction of Cr spins is symmetric. The Fermi level ($E_F$), indicated by a vertical line, is taken as the reference energy.}
    \label{fig:edos}
\end{figure}

The resulting projected atomic and orbital electronic densities of states in CrN without defects, CrN$_{1-x}$ and Cr$_{1-x}$N, which are shown in Fig.~\ref{fig:edos}, clearly indicate that the electronic structures of these systems are dominated by Cr-$3d$ and N-$2p$ states in a broad energy range, with the former having a significant contribution. The overlap of these states is especially pronounced at the top of the valence band, reflecting strong hybridization of Cr-$3d$ and N-$2p$ states near the Fermi level, as well as over an energy range extending up to about 0.5 eV below $E_F$. In the conduction band, the spectral density of Cr-$3d$ electrons is much higher than that of N-$2p$ electrons. The calculated band gap of defect-free bilayer CrN is 0.38 eV, which is in good agreement with the experimentally determined indirect optical band gap of nearly stoichiometric samples deposited at temperatures below 500 °C. This value is also consistent with previously reported band gaps ranging from 0.2 to 0.5 eV for epitaxial CrN(001) films grown on MgO(001) substrates \cite{gall_1,gall_2}. The cation (V$_{\mathrm{Cr}}$) and anion (V$_{\mathrm{N}}$) vacancies notably affect the electronic structure of the CrN bilayer. Significantly reduced (0.07 eV) and vanished band gaps are observed in CrN$_{1-x}$ and Cr$_{1-x}$N, respectively.  A detailed examination of the electronic projected densities of states in CrN with V$_{\mathrm{N}}$/V$_{\mathrm{Cr}}$ together with the Bader charge analysis \cite{bader} allows us to identify the origin of these changes. The Bader charges of Cr and N ions in defect-free CrN amount to $+2.20e$ and $-2.20e$, respectively. They are lower than the nominal ionic charges of Cr$^{3+}$ and N$^{3-}$ due to the covalency effects that reduce the charge transfer between cations and anions. In the system with N-vacancies, the $3d$ bands arising from unsaturated bonds of Cr ions surrounding V$_{\mathrm{N}}$ appear at the top of the valence band, resulting in an upward shift of the Fermi level and a considerably decreased band gap. The effective charges of Cr ions adjacent to V$_{\mathrm{N}}$ decrease to $+1.70e$. Hence, CrN$_{1-x}$ with $x=3$ at. \% remains a semiconductor with a very small energy gap. A higher content of N vacancies would possibly convert the semiconducting bilayer CrN to a metallic or degenerate semiconducting state. The formation of vacancies in the cation sublattice induces additional states within the forbidden band gap, leading to its closure. The band gap states are filled by the $2p$ states of the N ions neighboring V$_{\mathrm{Cr}}$, which are hybridized with the $3d$ states of Cr ions being the second-nearest neighbors of the Cr vacancy. The valence charges of these Cr ions increase to $+2.53e$, indicating that an electron charge transfer takes place.  Charges are transferred to the nitrogen $ p$ band to compensate for holes created after Cr is removed from the insulating compound. Thus, the CrN$_{1-x}$ bilayer becomes a metallic system for $x=3$ at.\%. The strong dependence of the electronic structure of CrN on stoichiometry (N/Cr deficiency) and defect concentration affects the electronic and thermal transport properties of CrN thin films, including resistivity, electronic thermal conductivity, and the Seebeck coefficient, ultimately modifying the thermoelectric properties of this material.  

\begin{figure}[ht]
    \centering
    \includegraphics[width=0.9\linewidth]{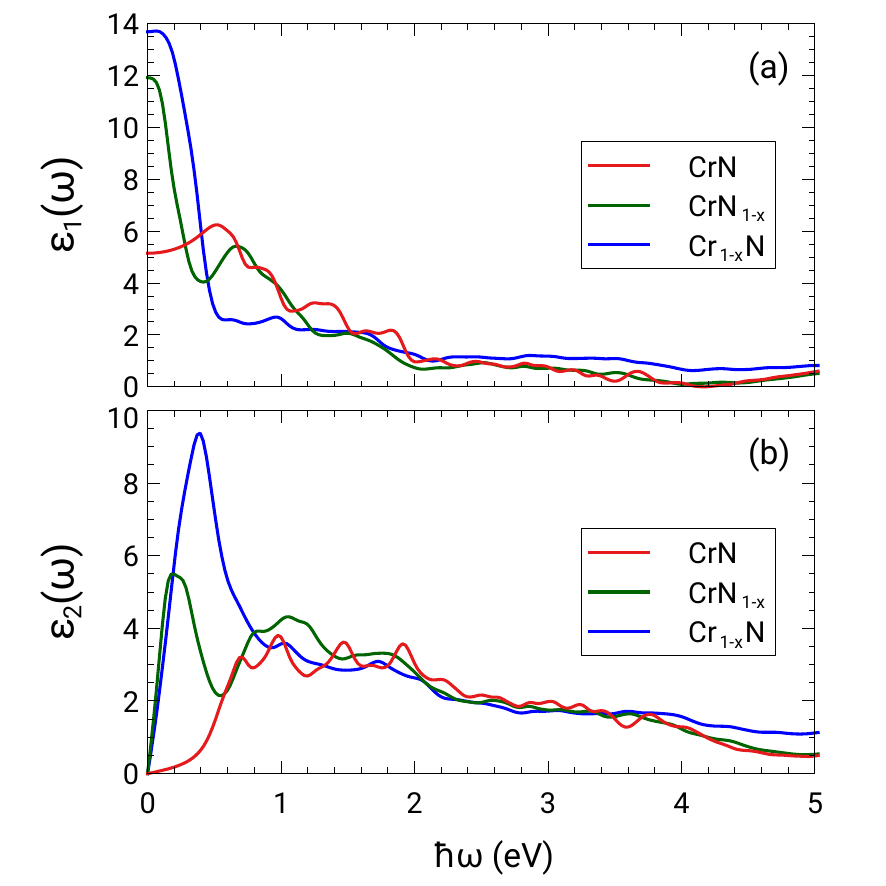}
    \caption{Photon energy dependent real $\varepsilon_1(\omega)$ and imaginary $\varepsilon_2(\omega)$ parts of dielectric functions calculated for CrN, CrN$_{1-x}$, and Cr$_{1-x}$N bilayers.}
    \label{fig:diel}
\end{figure}

Changes in the electronic structure of CrN, induced by the incorporated V$_{\mathrm{N}}$/V$_{\mathrm{Cr}}$ vacancies, are reflected in the linear response of CrN to electromagnetic radiation, which is characterized by the energy-dependent complex dielectric function $\varepsilon(\omega)=\varepsilon_1(\omega)+i \varepsilon_2(\omega)$. In general, the real $\varepsilon_1(\omega)$ and imaginary $\varepsilon_2(\omega)$ parts of $\varepsilon(\omega)$ describe the dispersion and absorption of the radiation in a given material, respectively.  The zero-frequency limit of the real part, $\varepsilon_1(\omega \rightarrow 0)$, corresponds to the electronic part of the static dielectric constant of a material $\varepsilon_{s}$. Figure \ref{fig:diel} shows the sensitivity of both components of $\varepsilon(\omega)$ to vacancies in the N or Cr sublattice for the low photon energy regime ($\hbar \omega < 2$ eV), while at higher photon energies the difference in dielectric functions between defect-free and defect-containing systems remains less pronounced. The $\varepsilon_{s}$ increases from 5.2 in CrN to 11.9 in CrN$_{1-x}$ and 13.7 in Cr$_{1-x}$ N with $x=3$ at.\% and remains, however, lower than the value of $\sim 20$ reported for CrN/MgO(001) \cite{gall_2}.  The much higher value of $\varepsilon_{s}$ deduced from the experimental spectra may indicate the existence of a larger number of delocalized charge carriers in much more non-stoichiometric samples that can easily be polarized, enhancing the value of $\varepsilon_{s}$. One should also note that experiments provide the static dielectric constant, \emph{i.e.,} the total dielectric screening from both the high-frequency response ($\varepsilon_{\infty}$)  and the vibrational (lattice) response. In contrast, most calculations are performed for static crystals (at zero temperature and with zero-point vibrations neglected), and hence yield values arising purely from electronic screening. 

\begin{figure}[h]
    \centering
    \includegraphics[width=0.75\linewidth]{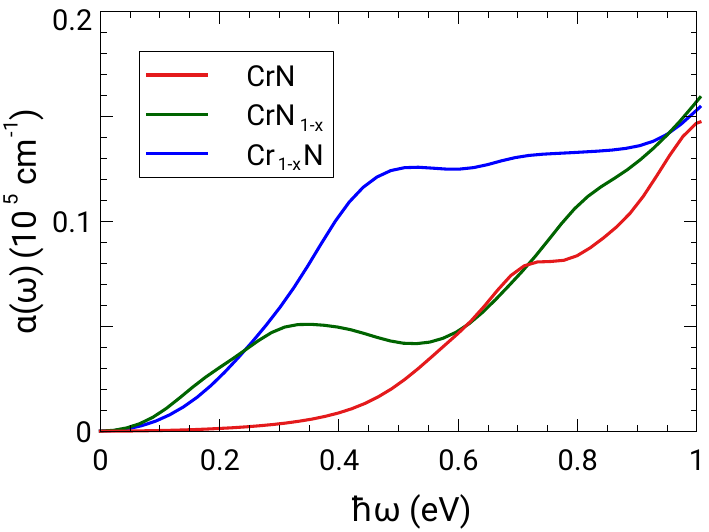}
    \caption{Photon energy dependent absorption coefficient $\alpha(\omega)$ calculated for CrN, CrN$_{1-x}$, and Cr$_{1-x}$N bilayers. Absorption coefficient is derived from real and imaginary parts of $\varepsilon(\omega)$ as  $\alpha(\omega)=\frac{\sqrt{2}\omega}{c}\left( \sqrt{\varepsilon_1^2(\omega)+\varepsilon_2^2(\omega)}-\varepsilon_1(\omega)\right)^{\frac{1}{2}}$, where $c$ is the speed of light.}
    \label{fig:alpha}
\end{figure}

The $\varepsilon_2(\omega)$ spectra clearly show the red shift of the absorption edge and a substantial reduction of the optical gap in CrN$_{1-x}$ as well as the absence of a gap in Cr$_{1-x}$N bilayers. This is also evidenced by the behavior of the absorption coefficient $\alpha(\omega)$ depicted in Fig.~\ref{fig:alpha}. This trend is consistent with the experimental part, where a decrease in the optical band gap (Fig. \ref{fig:Taucplot}, Table \ref{tab:Fit_N2Ar}), accompanied by a shift of the absorption edge (Fig. \ref{fig:eps12_N2Ar}), is observed for N-rich samples.

The $\varepsilon_1(\omega)$ and $\varepsilon_2(\omega)$ spectra show growth with $\omega \rightarrow 0$, reflecting the narrow-gap nature of CrN$_{1-x}$ and \emph{the metallization} of Cr$_{1-x}$N. The increase in the real part of the dielectric function at $\omega \rightarrow 0$  yields a pronounced increase in the electronic part of the static dielectric constant in CrN with nitrogen and chromium vacancies. The increase in absorption spectra at low energies, typical of metals (Drude peak), indicates the existence of delocalized charge carriers, regardless of the Mott/charge transfer insulator state of chromium nitride \cite{herwadkar}. These itinerant charge carriers (electrons and holes) arise from native point defects in CrN, such as vacancies in its cation/anion sublattice. 

\begin{figure}[h]
    \centering
    \includegraphics[width=1.0\linewidth]{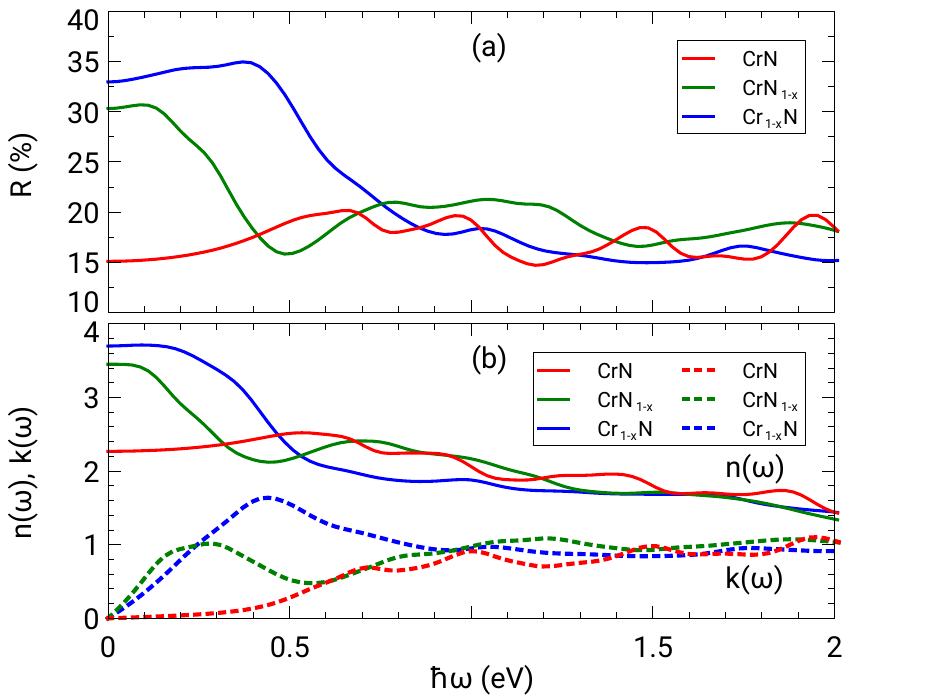}
    \caption{Photon energy dependent (a) reflectivity $R(\omega)$, (b) refractive index $n(\omega)$ and extinction coefficient $k(\omega)$ calculated for CrN, CrN$_{1-x}$, and Cr$_{1-x}$N bilayers. Solid and dashed lines correspond to $n(\omega)$ and $k(\omega)$, respectively.}
    \label{fig:Rnk}
\end{figure}

The real and imaginary components of $\varepsilon(\omega)$ can be used to derive reflectivity $R(\omega)$, index of refraction $n(\omega)$, and extinction coefficient $k(\omega)$ according to the following relations \cite{fox}:
\begin{eqnarray}
    R(\omega)= \left|\frac{\sqrt{\varepsilon(\omega)}-1}{\sqrt{\varepsilon(\omega)}+1}\right|^2 , \\ \nonumber
    n(\omega)=\sqrt{\frac{|\varepsilon(\omega)|+\varepsilon_1(\omega)}{2}} , \\ \nonumber
    k(\omega)=\sqrt{\frac{|\varepsilon(\omega)|-\varepsilon_1(\omega)}{2}}.
\end{eqnarray}
The calculated energy dependence of $R(\omega)$, $n(\omega)$, and $k(\omega)$ in CrN, CrN$_{1-x}$, and Cr$_{1-x}$N bilayers is shown in Fig.~\ref{fig:Rnk}. Similar to the behavior of the dielectric function, the zero-energy limit of the reflectivity and refractive index increases twice in CrN containing vacancies, and the absorption edge visible in the extinction coefficient diminishes (vanishes) for CrN$_{1-x}$ (Cr$_{1-x}$N). 

\section{Summary and conclusions}
In this work, we report a combined experimental and \textit{ab initio} approach to systematically investigate the structural, optical, and electronic properties of $\text{CrN}$ thin films deposited on fused silica and $\text{MgO}$ substrates over a wide temperature range from $400 \degree$C to $800 \degree$C. 

Structural characterization by XRD and AFM revealed a strong dependence of the film morphology on the substrate and deposition temperature. Cube-on-cube epitaxial growth was achieved on MgO, while the grain size increased systematically with increasing substrate temperature. This structural evolution is accompanied by significant changes in the films' electrical and optical properties.

We demonstrate that the electrical transport properties of CrN can be effectively tuned through nonstoichiometry controlled by the $\text{N}_2/\text{Ar}$ gas-flow ratio and substrate temperature. Nitrogen-rich deposition conditions promote $p$-type conductivity, with the lowest resistivity obtained at a substrate temperature of approximately 500 $\degree $C and an $\text{N}_2/\text{Ar}$ ratio of 4. At temperatures above 600 $\degree$C, the conductivity changes to $n$-type. The experimental results suggest that this behavior is associated with changes in the dominant native defects: nitrogen-rich growth favors Cr vacancies, which promote $p$-type transport, whereas elevated deposition temperatures favor nitrogen loss and the formation of N vacancies, resulting in $n$-type conduction.

The optical measurements further demonstrate a strong dependence of the dielectric response and optical absorption edge on film stoichiometry and deposition conditions. In particular, changes in the low-energy dielectric response correlate with observed variations in electrical resistivity, indicating a close relationship between defect-induced changes in the electronic structure and charge-carrier dynamics.

The experimental observations are strongly supported by the \textit{ab initio} calculations. The calculations show that Cr vacancies substantially modify the electronic structure and lead to closure of the band gap at the investigated vacancy concentration, whereas N vacancies strongly reduce the band gap. These results provide a microscopic explanation for the pronounced sensitivity of the electronic and optical properties of CrN to deviations from stoichiometry.

Overall, the results demonstrate that reactive magnetron sputtering provides an effective route for controlling the defect chemistry and, consequently, the electrical and optical properties of CrN thin films. The ability to tune the carrier type from $p$-type to $n$-type via deposition conditions makes CrN a promising material for developing functional multilayer structures and for potential thermoelectric and optoelectronic applications.

\section*{Declaration of competing interest}
The authors declare that they have no known competing financial interests or personal relationships that could have appeared to influence the work reported in this paper.

\section*{Acknowledgments}
This work was supported by the Czech Science Foundation (GACR, project No. 23-07228S); 
Ministry of Education, Youth and Sports of the Czech Republic (projects SENDISO - CZ.02.01.01/00/22\_008/0004596, QM4ST - CZ.02.01.01/00/22\_008/0004572, and the e-INFRA CZ - ID:90254). 
The Interdisciplinary Center of Mathematical and Computational Modeling (ICM), Warsaw University, Poland, is acknowledged for providing computing facilities for some of the calculations presented here. 

\section*{Data availability}
Data will be made available on request.

\section*{Declaration of generative AI and AI-assisted technologies in the manuscript preparation process}
During the preparation of this work, the author(s) used Grammarly and ChatGPT AI tool for grammar correction and rephrasing of the text. The author(s) reviewed and edited the output as needed and take full responsibility for the content of the published article.









\printcredits

\bibliographystyle{elsarticle-num}

\bibliography{references}



\end{document}